\documentclass[aps,pre, reprint, twocolumn, superscriptaddress,nofootinbib]{revtex4-2}

\usepackage{graphicx}
\usepackage{xcolor}
\usepackage{amsmath,amssymb,bm}
\usepackage{mathtools}
\usepackage{dsfont}
\usepackage{hyperref}
\usepackage{siunitx}
\usepackage{booktabs}
\usepackage{enumitem}

\newcommand{\abs}[1]{\left| #1 \right|}

\newcommand{\kpeak}{k_{\mathrm{peak}}}
\newcommand{\Kmax}{K_{\max}}
\newcommand{\eps}{\varepsilon}
\newcommand{\dt}{\Delta t}

\newif\ifshowchanges
\ifshowchanges
  \DeclareRobustCommand{\HLF}[1]{{\color{red}#1}}
  \DeclareRobustCommand{\HLS}[1]{{\color{blue}#1}}
  \DeclareRobustCommand{\HLT}[1]{{\color{green!45!black}#1}}
\else
  \DeclareRobustCommand{\HLF}[1]{#1}
  \DeclareRobustCommand{\HLS}[1]{#1}
  \DeclareRobustCommand{\HLT}[1]{#1}
\fi

\begin{document}

\title{Extreme-value statistics of curl-of-vorticity precursor peaks in perturbed Taylor--Green vortex turbulence}

\author{Satori Tsuzuki}
\affiliation{Research Center for Advanced Science and Technology, The University of Tokyo}

\date{\today}

\begin{abstract}
Precursor peaks in the wavenumber $k_{\mathrm{peak}}(t)$ maximizing the curl-of-vorticity spectrum have been observed to precede the dissipation peak in decaying turbulence. 
Because small perturbations in the initial condition can shift peak times, the associated lead time should be characterized statistically. 
We perform a pseudospectral DNS ensemble of $N_s=1000$ perturbed Taylor--Green vortex realizations at $N=256^3$ and $\nu=10^{-3}$. 
For each run we extract $k_{\mathrm{peak}}(t)$, several definitions of the precursor time $t_k$, the dissipation-peak time $t_\varepsilon$, and run-wise extrema including $K_{\max}=\max_t k_{\mathrm{peak}}(t)$ and $M_{\max}=\max_t\max_k \mathcal{C}(k,t)$, where $\mathcal{C}(k,t)$ is the isotropic curl-of-vorticity spectrum. The distribution of $\Delta t_{\varepsilon,k}=t_\varepsilon-t_k$ shows that the precursor typically leads, while rare lagging realizations occur and are strongly conditioned on $K_{\max}$. 
\HLT{Using peaks-over-threshold extreme-value theory, we fit generalized Pareto models to the right tails of $X=-\Delta t_{\varepsilon,k}$ and $M_{\max}$; the negative shape estimates are consistent with effective bounded tails under the present finite-resolution sampling protocol and provide protocol-dependent endpoint estimates.} 
\HLT{Finally, $M_{\max}$ correlates strongly with $\varepsilon_{\max}$ and ensemble cross-correlations reveal a reproducible phase offset, consistent with an empirical association between high-curvature activity and dissipation bursts.}
\end{abstract}

\maketitle

\section{Introduction}\label{sec:intro}
Intermittent dissipation bursts are a defining feature of turbulence and motivate efforts to characterize not only their intensity but also their timing. In high-Reynolds-number flows, nonlinear advection transfers energy across a wide range of scales, intermittency emerges, and the dynamics often organizes into coherent vortical structures embedded in seemingly irregular motion. In the continuum regime, this evolution is governed by the deterministic incompressible Navier--Stokes equations; accordingly, a pseudospectral DNS produces a unique trajectory once the initial condition and physical parameters are specified. At the same time, turbulence is notoriously sensitive to small changes: tiny perturbations in the initial state or in modeling and discretization details can shift the occurrence times of extreme-gradient events, including the dissipation peak. Therefore, the timing of such events should be assessed statistically across an ensemble of perturbed realizations, and the practical performance of any precursor must be quantified in terms of reliability rather than by a single deterministic ordering. This viewpoint sets the stage for recent observations of spectral precursor peaks that tend to precede the dissipation maximum in decaying turbulence.

\HLF{The Taylor--Green vortex (TGV) has long served as a canonical configuration for studying the generation of small scales from a simple, highly symmetric initial condition~\cite{TaylorGreen1937,Brachet1983JFM}. Classical direct numerical simulations of the TGV established it as a controlled setting in which vortex stretching, symmetry breaking, spectral transfer, and the emergence of small-scale turbulent structure can be examined. For this reason, the TGV is also widely used as a benchmark for numerical methods for transitional and decaying turbulence. In the present work, we use the TGV in this controlled sense: the objective is not to claim universality of a precursor relation, but to test, within a well-defined canonical flow, how a previously observed spectral timing relation behaves under a large ensemble of small perturbations.}

A recently identified precursor is based on the curl-of-vorticity spectrum.
For the unperturbed Taylor--Green vortex (TGV), Tsuzuki~\cite{Tsuzuki2026PRFluids} reported that the time at which the peak wavenumber $k_{\mathrm{peak}}(t)$ attains its maximum precedes the time of maximum viscous dissipation.
A subsequent study~\cite{Tsuzuki2026arXiv} showed that this ordering persists for moderate viscosities (e.g., $\nu=10^{-3}$ or $2.5\times10^{-4}$) and several initial conditions, while it can be weakened or lost in highly viscous, diffusion-dominated cases (e.g., $\nu=10^{-2}$).
\HLF{For the unperturbed TGV at $\nu=2.5\times10^{-4}$, Ref.~\cite{Tsuzuki2026arXiv} used a companion $1024^3$ simulation to avoid cutoff-proximate peak locking in the $k^4$-weighted (curl-of-vorticity) spectrum.} These findings suggest that the precursor is robust yet problem-dependent. This motivates the central question addressed here: How reliable is the observed lead time under small but unavoidable perturbations?

The central goal of this paper is to quantify the uncertainty of the precursor timing in the canonical TGV when the initial condition is subjected to small random perturbations representing experimental or numerical uncertainties.
We focus on two complementary aspects:
(i) the \emph{probabilistic reliability} of the precursor as measured by the distribution of $\dt_{\eps,k}=t_\eps-t_k$ and related conditional probabilities, and
(ii) the \emph{extreme-tail} behavior of large lags, modeled using peaks-over-threshold (POT) extreme-value theory.
We translate a deterministic spectral precursor into a probabilistic reliability problem under unavoidable initial-condition uncertainty, and quantify worst-case failures via POT extreme-value theory.

\paragraph*{Extreme-value theory background.}
Extreme-value theory (EVT) provides a probabilistic framework for modeling the tails of distributions and the statistics of unusually large (or small) observations.
Let $\{X_i\}_{i=1}^n$ be independent and identically distributed (i.i.d.) random variables with cumulative distribution function (CDF) $F$, and let $Z_n=\max_{1\le i\le n}X_i$ denote the sample maximum.
The Fisher--Tippett--Gnedenko theorem states that if there exist normalizing constants $a_n>0$ and $b_n$ such that $(Z_n-b_n)/a_n$ converges in distribution to a nondegenerate limit, then that limit must belong to one of three types (Gumbel, Fr\'echet, or Weibull) and can be written in the unified generalized extreme-value (GEV) form~\cite{FisherTippett1928,Gnedenko1943}
\begin{align}
G(z;\mu,\sigma,\xi)&=\exp\!\left\{-\left[1+\xi\frac{z-\mu}{\sigma}\right]^{-1/\xi}\right\},  \nonumber \\ 
&\qquad 1+\xi\frac{z-\mu}{\sigma}>0,
\label{eq:gev}
\end{align}
where $\mu\in\mathbb{R}$ is a location parameter, $\sigma>0$ is a scale parameter, and $\xi$ is the shape parameter (with the Gumbel case recovered as $\xi\to 0$).
Distributions $F$ for which such a limit exists are said to belong to the max-domain of attraction of a GEV law, and this condition underpins the block-maxima approach: one partitions a record into blocks, takes the maximum in each block, and fits a GEV model to the resulting sequence of block maxima.

A practical limitation of block maxima is that the number of maxima available for inference equals the number of blocks, which can be small when the record length is limited or when large blocks are required to justify approximate independence.
To make more efficient use of tail information, EVT often adopts the peaks-over-threshold (POT) formulation, which models exceedances over a high threshold.

\paragraph*{Peaks-over-threshold and generalized Pareto tails.}
In POT, one selects a high threshold $u$ and considers the conditional excess distribution of $Y=X-u$ given $X>u$,
\begin{align}
F_u(y)&=\Pr(X-u\le y\,|\,X>u) \nonumber \\
&=\frac{F(u+y)-F(u)}{1-F(u)},~~ y\ge 0.
\label{eq:excess_cdf}
\end{align}
The Pickands--Balkema--de~Haan theorem states that if $F$ lies in a max-domain of attraction (the same broad condition that justifies GEV limits), then for sufficiently high $u$ the excess distribution $F_u$ is well approximated by a generalized Pareto distribution (GPD)~\cite{Pickands1975,BalkemaDeHaan1974}
\begin{align}
H(y;\xi, \beta) &=
\begin{cases}
1-\left(1+\xi \dfrac{y}{\beta}\right)^{-1/\xi}, & \xi\neq 0,\\[6pt]
1-\exp\!\left(-\dfrac{y}{\beta}\right), & \xi=0,
\end{cases}
\nonumber \\
&~~~~\quad y\ge 0,\ \ 1+\xi \dfrac{y}{\beta}>0,
\label{eq:gpd}
\end{align}
where $\beta>0$ is a scale parameter and $\xi$ is the shape parameter.
\HLF{The parameter $\xi$ controls the tail type: $\xi>0$ corresponds to heavy-tailed behavior, $\xi=0$ to exponential-type tails, and $\xi<0$ to bounded upper tails.}
In the bounded case $\xi<0$, the fitted model implies a finite upper endpoint for $X$,
\begin{equation}
x_{\mathrm{end}} \equiv u - \frac{\beta}{\xi},
\qquad (\xi<0),
\label{eq:endpoint}
\end{equation}
which is particularly useful when estimating worst-case scenarios (e.g., the largest plausible lag).
More generally, POT enables estimation of high quantiles (return levels) using exceedance probabilities and the fitted GPD parameters.

In this work we apply POT to run-wise observables obtained from an ensemble of perturbed DNS realizations.
Specifically, we analyze the right tails of $X=-\Delta t_{\varepsilon,k}$ (rare lagging cases) and of $M_{\max}$, using high-quantile thresholds to maximize tail-information usage while retaining sufficient exceedances for stable inference (see Secs.~\ref{sec:methods}--\ref{sec:evt} and Appendix~\ref{app:pot_diagnostics}).

In related works, EVT and extreme/rare-event statistics have been increasingly used in fluid mechanics to connect intermittent bursts and regime changes to the tail behavior of physically meaningful observables. In transitional wall-bounded turbulence, Goldenfeld \emph{et al.} linked the finite lifetime of turbulent transients to \emph{extreme} fluctuations that drive decay, providing a statistical rationale for observed lifetime scalings \cite{Goldenfeld2010PRE}.  Building on the idea that changes in stability should imprint themselves on extremes, Faranda \emph{et al.} applied EVT-based diagnostics to detect global stability thresholds in transitional plane Couette flow \cite{Faranda2014CSF}.  In pipe flow, Nemoto and Alexakis directly examined whether large-amplitude bursts in turbulence intensity can trigger turbulent decay, quantifying how extreme events relate to decay-time statistics \cite{NemotoAlexakis2021JFM}; a broader perspective on how such extreme events manifest across transitional turbulence is summarized by Gom\'e \emph{et al.} \cite{GomeTuckermanBarkley2022PRSA}. For fully developed Navier--Stokes turbulence, Yeung \emph{et al.} characterized extreme events in DNS and their implications for small-scale intermittency \cite{Yeung2015PNAS}, while Saw \emph{et al.} provided an experimental characterization of extreme inertial-dissipation events in a turbulent swirling flow \cite{Saw2016NatCommun}.  Focusing on tail statistics of small-scale gradients, Buaria \emph{et al.} analyzed extreme velocity-gradient fluctuations \cite{Buaria2019NJP} and later showed how the most intense events can self-attenuate through local dynamical mechanisms \cite{Buaria2020NatCommun}.  Complementing classical block-maxima or peak-over-threshold fitting, Farazmand and Sapsis introduced a variational strategy to probe and uncover dynamical precursors of extremes in turbulent systems \cite{FarazmandSapsis2017SciAdv}, and Sapsis reviewed a wider toolbox for extreme-event statistics and prediction in fluid flows and waves \cite{Sapsis2021ARFM}.

\HLF{While Refs.~\cite{Tsuzuki2026PRFluids, Tsuzuki2026arXiv} focused on deterministic ordering and robustness of precursor peaks in selected realizations, the present work treats the lead--lag time as a run-wise random variable induced by initial-condition uncertainty. The main contribution is therefore not a new deterministic precursor mechanism, but a probabilistic reliability assessment of an observed spectral timing relation over $N_s=1000$ perturbed TGV realizations, including conditional failure probabilities and peaks-over-threshold estimates for rare lagging events.} Building on the above studies, this study advances the field in three essential directions. First, rather than analyzing isolated extreme bursts or transition events, it formulates the \emph{lead-lag relation} between a spectral precursor $k_{\mathrm{peak}}(t)$ and the dissipation peak as a statistically well-defined random variable and quantifies it over a large DNS ensemble ($N_s=1000$), thereby moving from deterministic observation to probabilistic characterization. Second, by applying a peaks-over-threshold extreme-value framework to the right tails of $X=-\Delta t_{\varepsilon,k}$ and of the run-wise maximum $M_{\max}$, \HLT{it provides effective bounded-tail estimates and protocol-dependent worst-case lag estimates}---an aspect that has not been explicitly quantified in prior turbulence-EVT studies. Third, the demonstrated strong correlation between $M_{\max}$ and $\varepsilon_{\max}$, together with reproducible ensemble phase offsets, \HLT{provides statistical evidence for an empirical association between high-curvature spectral activity and dissipation bursts, rather than a causal derivation of the underlying Navier--Stokes mechanism.} In this way, the study integrates precursor detection, ensemble uncertainty quantification, and EVT-based tail modeling into a unified framework for assessing predictability and risk of extreme dissipation events in decaying turbulence.

The remainder of this paper is organized as follows.
Section~\ref{sec:methods} describes the numerical methodology, ensemble construction, and definitions of the spectral observables and timing metrics.
Section~\ref{sec:evt} outlines the peaks-over-threshold extreme-value framework and the associated inference procedure.
Section~\ref{sec:results} presents the ensemble statistics and POT results.
Section~\ref{sec:discussion} discusses the physical interpretation, regime structure, and limitations of the analysis.
Finally, Section~\ref{sec:conclusions} summarizes the main findings and outlines directions for future work.

\section{Numerical method and datasets}
\label{sec:methods}

\subsection*{Overview of DNS}
\label{sec:dns}

We solve the incompressible Navier--Stokes equations in a periodic cube,
\begin{align}
\partial_t \bm{u} + \bm{u}\cdot\nabla \bm{u} &= -\nabla p + \nu \Delta \bm{u}, \\
\nabla\cdot \bm{u} &= 0,
\end{align}
on $[0,2\pi]^3$ using a Fourier pseudospectral method with the $2/3$ de-aliasing rule.
Time integration is performed with a classical four-stage fourth-order Runge--Kutta method (RK4) together with an integrating-factor treatment of the viscous term, as in Refs.~\cite{Tsuzuki2026PRFluids,Tsuzuki2026arXiv}.
Unless stated otherwise, the reference ensemble uses $N=256^3$ grid points, viscosity $\nu=10^{-3}$, and a fixed time step $\Delta t=10^{-3}$.

\HLF{The triply periodic geometry makes a Fourier pseudospectral method natural, because spatial derivatives are evaluated spectrally and the incompressibility constraint can be enforced mode by mode by projection onto divergence-free Fourier modes. The nonlinear term is evaluated in physical space and transformed back to Fourier space; the standard $2/3$ de-aliasing rule is used to remove aliasing errors generated by this quadratic convolution. The linear viscous term is treated with an integrating factor, which accounts for the diffusive decay of each Fourier mode over a time step, while the remaining nonlinear dynamics are advanced with a classical fourth-order Runge--Kutta scheme. This combination is standard for DNS of incompressible periodic turbulence and is well suited to the spectral diagnostics used below.}

We store shell-averaged spectra at a cadence of $50$ RK steps, corresponding to an output 
interval $\Delta t_{\mathrm{out}} = 0.05$ in nondimensional units.
All characteristic times reported below (e.g., $t_k$, $t_{\varepsilon}$, and $t_M$) are extracted from these
discretely sampled outputs and are therefore quantized in steps of $\Delta t_{\mathrm{out}}$.
Conservatively, we assign an uncertainty $|\delta t| \le \Delta t_{\mathrm{out}}$ to each extracted event time,
so that timing differences such as $\Delta t_{\varepsilon,k} \equiv t_{\varepsilon} - t_k$ have an uncertainty
bounded by $|\delta \Delta t_{\varepsilon,k}| \le 2\Delta t_{\mathrm{out}}$.
This choice follows Ref.~\cite{Tsuzuki2026arXiv}, which used the same output interval under the same pseudo-spectral DNS protocol and
showed that, in adequately resolved $\nu=10^{-3}$ runs (passing spike inspection), the smallest separation between
the peak-scale time and the dissipation-peak time was $|\Delta t_{\varepsilon,k}| = 0.40 \simeq 8\,\Delta t_{\mathrm{out}}$,
well above the cadence-induced discretization bound.
Using the same $\Delta t_{\mathrm{out}}$ here therefore enables direct comparison with Ref.~\cite{Tsuzuki2026arXiv} while providing
adequate temporal resolution for the present parameter regime. 
Accordingly, when reporting lag-event probabilities we adopt a tolerance $\delta = 2\Delta t_{\mathrm{out}}$ and classify a realization as a ``lag event'' only when $\Delta t_{\varepsilon,k} < -\delta$.

\begin{table*}[t]
\caption{DNS and dataset parameters for the main ensemble analyzed in this work.}
\label{tab:simparams}
\begin{ruledtabular}
\begin{tabular}{ll}
Domain & $[0,2\pi]^3$ (periodic) \\
Grid & $N=256^3$ \\
De-aliasing & $2/3$ rule \\
Viscosity (reference) & $\nu=10^{-3}$ \\
Time integration & RK4 with integrating factor (viscous term) \\
Time step & $\Delta t = 10^{-3}$ \\
Spectra sampling & every 50 steps ($\Delta t_{\mathrm{out}}=0.05$) \\
End time & $t_{\mathrm{end}}=20$ \\
Ensemble size & $N_s=1000$ independent perturbation seeds \\
Initial total kinetic energy & $E_0=0.125$ \\
Perturbation weight & $w=0.05$ (see Sec.~\ref{sec:pert}) \\
Perturbation bandwidth & $k \le 3$ (low-$k$ random field) \\
\end{tabular}
\end{ruledtabular}
\end{table*}

\subsection*{Initial condition perturbations}\label{sec:pert}
The baseline initial condition is the Taylor--Green vortex, originally introduced by Taylor and Green~\cite{TaylorGreen1937} and widely used as a canonical decaying-turbulence configuration.
Following the approach of Refs.~\cite{Tsuzuki2026PRFluids,Tsuzuki2026arXiv}, we generate a family of perturbed initial conditions of the form
\begin{equation}
\bm{u}(\bm{x},0) = \bm{u}_{\mathrm{TGV}}(\bm{x}) + w\, \bm{u}_{\mathrm{rand}}(\bm{x}),
\end{equation}
where $w=0.05$ and $\bm{u}_{\mathrm{rand}}$ is a divergence-free random field whose Fourier support is limited to low wavenumbers ($k \le 3$).
For each seed, the perturbation phases are randomized and the full initial condition is rescaled such that the total kinetic energy is fixed at $E_0=0.125$.
This construction isolates the effect of small-scale-agnostic initial uncertainty while keeping the gross energy level identical across the ensemble.

\subsection*{Spectral observables and timing definitions}\label{sec:obs}
We define vorticity $\bm{\omega}=\nabla\times \bm{u}$ and the curl of vorticity
\begin{equation}
\bm{\chi} \equiv \nabla\times \bm{\omega} = \nabla\times(\nabla\times\bm{u}) = -\Delta \bm{u},
\end{equation}
where the last identity uses incompressibility.
Consequently, the shell-averaged spectrum of $\abs{\bm{\chi}}^2$ is proportional to $k^4 E(k)$ for a solenoidal velocity field, and thus accentuates the development of small-scale content~\cite{Tsuzuki2026arXiv}.

Let $\mathcal{C}(k,t)$ denote the isotropic (shell-averaged) spectrum of $\abs{\bm{\chi}}^2$ at time $t$.
We define the \emph{peak wavenumber}
\begin{equation}
\kpeak(t) \equiv \arg\max_k \, \mathcal{C}(k,t),
\end{equation}
and its run-wise maximum
\begin{equation}
\Kmax \equiv \max_t \kpeak(t).
\end{equation}
The maximum amplitude of the curl-of-vorticity spectrum is
\begin{equation}
M(t) \equiv \max_k \mathcal{C}(k,t), \qquad M_{\max} \equiv \max_t M(t),
\end{equation}
with peak time $t_M=\arg\max_t M(t)$.

We also compute the viscous dissipation rate $\eps(t)$ (from the DNS outputs) and define the dissipation peak time
\begin{equation}
t_\eps \equiv \arg\max_t \eps(t), \qquad \eps_{\max} \equiv \max_t \eps(t).
\end{equation}

To extract precursor timing, we use three operational definitions of the time $t_k$ associated with the $\kpeak(t)$ evolution:
\begin{enumerate}[label=(\roman*), leftmargin=2.2em]
\item \emph{First hitting time} $t_{k,\mathrm{first}}$: the earliest time at which $\kpeak(t)$ reaches $\Kmax$.
\item \emph{Last hitting time} $t_{k,\mathrm{last}}$: the latest time at which $\kpeak(t)$ equals $\Kmax$.
\item \emph{Fractional time} $t_{k,s}$: the earliest time at which $\kpeak(t)\ge s\,\Kmax$ for a prescribed fraction $s$; in this work $s=0.95$. We write $t_{k,95}$ to indicate the 95\% level, using percentage notation for brevity.
\end{enumerate}

The precursor lead/lag time is then
\begin{equation}
\dt_{\eps,k} \equiv t_\eps - t_k,
\end{equation}
so that $\dt_{\eps,k}>0$ indicates a lead (precursor behavior) and $\dt_{\eps,k}<0$ indicates a lag.
In addition, we define the time offset between the dissipation-peak time and the peak time of the maximum curl-of-vorticity amplitude, denoted by $\Delta t_{\varepsilon,M}$, as
\begin{equation}
\Delta t_{\varepsilon,M} \equiv t_{\varepsilon} - t_{M}.
\end{equation}

We also define the ``plateau duration'' of the maximum-wavenumber state,
\begin{equation}
T_{\mathrm{plat}} \equiv t_{k,\mathrm{last}} - t_{k,\mathrm{first}},
\end{equation}
which quantifies how long $\kpeak(t)=\Kmax$ persists. 

\subsection*{Spike inspection}
\label{sec:spike}

Because $\mathcal{C}(k,t)$ emphasizes high wavenumbers, insufficient spatial resolution can cause spurious behavior in $\kpeak(t)$, in particular a ``cutoff-proximate peak locking'' where $\kpeak(t)$ sticks near the maximum resolved wavenumber.
To prevent contamination of the precursor statistics by such numerical artifacts, we apply a spike inspection procedure.
In brief, a run is flagged if $\kpeak(t)$ resides within a prescribed fraction (here $95\%$) of the de-aliased isotropic cutoff wavenumber for any portion of the evolution, or if associated diagnostic conditions indicate a nonphysical high-$k$ spike.
All results reported for the reference case ($N=256^3$, $\nu=10^{-3}$) pass this inspection.

\paragraph*{Spectral-resolution diagnostic.}
In addition to the spike inspection described above, we quantify the small-scale
resolution by monitoring the standard pseudospectral indicator $k_{\max}\eta(t)$.
Here $\eta(t)=(\nu^{3}/\varepsilon(t))^{1/4}$ is the Kolmogorov length scale, and
the dissipation rate is computed spectrally from the energy spectrum as
$\varepsilon(t)=2\nu\sum_{k} k^{2}E(k,t)$.
We report $k_{\max}$ at the Nyquist wavenumber ($k_{\max}=N/2$) and also the
effective maximum wavenumber after the $2/3$ de-aliasing rule
($k_{\rm cut}=N/3$, so that $k_{\rm cut}\eta=(2/3)k_{\max}\eta$).
For the resolved reference ensemble ($N=256^3$, $\nu=10^{-3}$, $N_s=1000$),
the minimum value over time, $\min_t(k_{\max}\eta)$, is narrowly distributed
across seeds: $2.165 \le \min_t(k_{\max}\eta) \le 2.194$
(median $2.180$, IQR $[2.177,\,2.183]$).
The corresponding de-aliased measure satisfies
$1.443 \le \min_t(k_{\rm cut}\eta) \le 1.463$ (median $1.453$).
In every run the minimum $k_{\max}\eta$ is attained at the dissipation-peak time $t_{\varepsilon}=\arg\max_t \varepsilon(t)$, i.e., $k_{\max}\eta(t_{\varepsilon})=\min_t(k_{\max}\eta)$.
A representative time series of $R_{\lambda}(t)$ and $k_{\max}\eta(t)$ is provided in Appendix~\ref{app:kmaxeta} (Fig.~\ref{fig:fig6}).
Unless otherwise stated, all numerical diagnostics, including spike inspection, spectral normalization,
shell averaging, and peak-time extraction, follow the conventions established in Ref.~\cite{Tsuzuki2026arXiv}.
Table~\ref{tab:simparams} summarizes the DNS setup and dataset parameters for the main ensemble analyzed in this work.
\HLF{The reliability of the reference ensemble is supported by three checks. First, all $N_s=1000$ realizations at $N=256^3$ and $\nu=10^{-3}$ pass the spike inspection designed to detect cutoff-proximate locking of $k_{\mathrm{peak}}(t)$. Second, the standard resolution indicator $k_{\max}\eta(t)$ remains comfortably above unity, and its minimum is attained at the dissipation peak in every run; in de-aliased units, $\min_t(k_{\rm cut}\eta)$ lies in $[1.443,1.463]$ across the ensemble. Third, the same diagnostics reject the lower-viscosity datasets, where $k_{\mathrm{peak}}(t)$ locks near the de-aliased cutoff. Thus, the statistical and EVT analyses are applied only to the resolved reference ensemble and not to spectra contaminated by high-wavenumber cutoff effects.}
The present study differs only in the ensemble construction and the subsequent extreme-value analysis.

\section{Extreme-value statistics method}\label{sec:evt}
The objective of the present analysis is to quantify not only the typical precursor lead time but also rare realizations in which the precursor fails, i.e., exhibits a substantial lag relative to the dissipation peak.
To this end, we model the upper tail of a scalar run-wise quantity $X$ using the peaks-over-threshold (POT) framework.
In the present context, $X$ is primarily taken to be the lag magnitude
\begin{equation}
X = -\Delta t_{\varepsilon,k},
\end{equation}
so that large values of $X$ correspond to strong lagging cases.
We also apply POT to $M_{\max}$ in order to characterize the extreme amplitude of the curvature proxy.

\subsection*{Run-wise sampling and independence}
Each realization in our ensemble corresponds to a distinct randomly perturbed initial condition.
For a fixed seed, the DNS evolution is deterministic; however, across seeds the extracted run-wise observables
(e.g., $\Delta t_{\varepsilon,k}$ and $M_{\max}$)
form a collection of independent samples under the assumption that the perturbations are independently drawn.
Thus, for the purpose of extreme-value modeling, we treat
\begin{equation}
X_1,\dots,X_{N_s}
\end{equation}
as approximately independent and identically distributed (i.i.d.) samples from an underlying distribution $F$.
\HLT{The independence assumption used here is a Monte-Carlo independence assumption with respect to the independently drawn perturbation seeds. It does not mean that the realizations explore unrelated regions of state space; all trajectories remain conditioned on the same TGV base flow, viscosity, resolution, and perturbation protocol. Thus, the fitted tail model describes the distribution induced by this perturbation ensemble. Because only one scalar observable is extracted from each deterministic run, the usual temporal clustering issue in POT analysis of time series is avoided, but the results should still be interpreted as ensemble-conditional rather than universal.}

\subsection*{Peaks-over-threshold formulation}
Given i.i.d. samples $X_1,\dots,X_n$ with cumulative distribution function $F$, the conditional excess distribution above a high threshold $u$ is defined as
\begin{equation}
F_u(y) \equiv \mathbb{P}(X-u \le y \mid X>u), 
\qquad y\ge 0.
\end{equation}
The Pickands--Balkema--de Haan theorem (see Appendix~\ref{app:gpd_excess}) states that if $F$ lies in a suitable max-domain of attraction, then for sufficiently high $u$ the excess distribution $F_u$ is well approximated by a generalized Pareto distribution (GPD).

The GPD is given by
\begin{align}
H(y;\xi, \beta) &=
\begin{cases}
1-\left(1+\xi \dfrac{y}{\beta}\right)^{-1/\xi}, & \xi\neq 0,\\[6pt]
1-\exp\!\left(-\dfrac{y}{\beta}\right), & \xi=0,
\end{cases}
\nonumber \\
&~~~~\quad y\ge 0,\ \ 1+\xi \dfrac{y}{\beta}>0,
\label{eq:gpd_repost}
\end{align}
where $\xi$ is the shape parameter and $\beta>0$ is the scale parameter.
\HLF{Equation~(\ref{eq:gpd_repost}) is a restatement of Eq.~(\ref{eq:gpd}).}
The sign of $\xi$ determines the tail class:
$\xi>0$ corresponds to heavy-tailed behavior,
$\xi=0$ to exponential-type decay,
and $\xi<0$ to a bounded upper tail.
In the latter case, the fitted model implies a finite upper endpoint
\begin{equation}
x_{\mathrm{end}} \equiv u - \frac{\beta}{\xi},
\qquad (\xi<0),
\end{equation}
\HLT{which provides a protocol-dependent estimate of the largest plausible extreme event under the fitted finite-resolution model.}

\subsection*{Threshold choice and inference}
In practice, we select a high empirical quantile threshold $u$, taken here as the $q=0.9$ quantile of the sample unless otherwise noted.
This choice represents a compromise: the threshold must be high enough for the GPD approximation to be valid, yet low enough to retain a sufficient number of exceedances for stable parameter estimation. For $N_s=1000$ and a $q = 0.9$ threshold, this typically yields $\mathcal{O}(10^{2})$ exceedances (here $n_{u} \approx$ 86--100 depending on ties; see Table~\ref{tab:potfits}).

We fit the GPD parameters $(\xi,\beta)$ to the exceedances
\begin{equation}
Y_i = X_i - u, \qquad X_i > u,
\end{equation}
by maximum likelihood, fixing the GPD location parameter at zero.
We report the fitted threshold $u$, the estimated shape and scale parameters, and, when $\xi<0$, the implied upper endpoint.
Sensitivity to the threshold is examined by varying $q$ in a moderate range and confirming qualitative robustness of the inferred tail class.
\HLT{Because $t_k$ and $t_\varepsilon$ are extracted from outputs separated by $\Delta t_{\mathrm{out}}=0.05$, the lag variable is discrete with a resolution-limited uncertainty of order $2\Delta t_{\mathrm{out}}$. The GPD fits should therefore be interpreted as continuous approximations to the tail of a discretely sampled observable. Endpoint estimates are not meaningful below this timing resolution, and the lag-event tolerance $\delta=2\Delta t_{\mathrm{out}}$ is used to avoid classifying cadence-level differences as genuine failures.}

\section{Results}\label{sec:results}
We first analyze the statistically resolved reference ensemble consisting of $N_s=1000$ perturbed realizations at $N=256^3$ and $\nu=10^{-3}$.
All runs pass the spike inspection and satisfy the resolution diagnostic discussed in Appendix~\ref{app:kmaxeta}, ensuring that the spectral peak statistics are not contaminated by cutoff locking.

\subsection{Representative realizations and precursor definitions}
\begin{figure*}[t]
  \centering
  \includegraphics[width=0.65\linewidth]{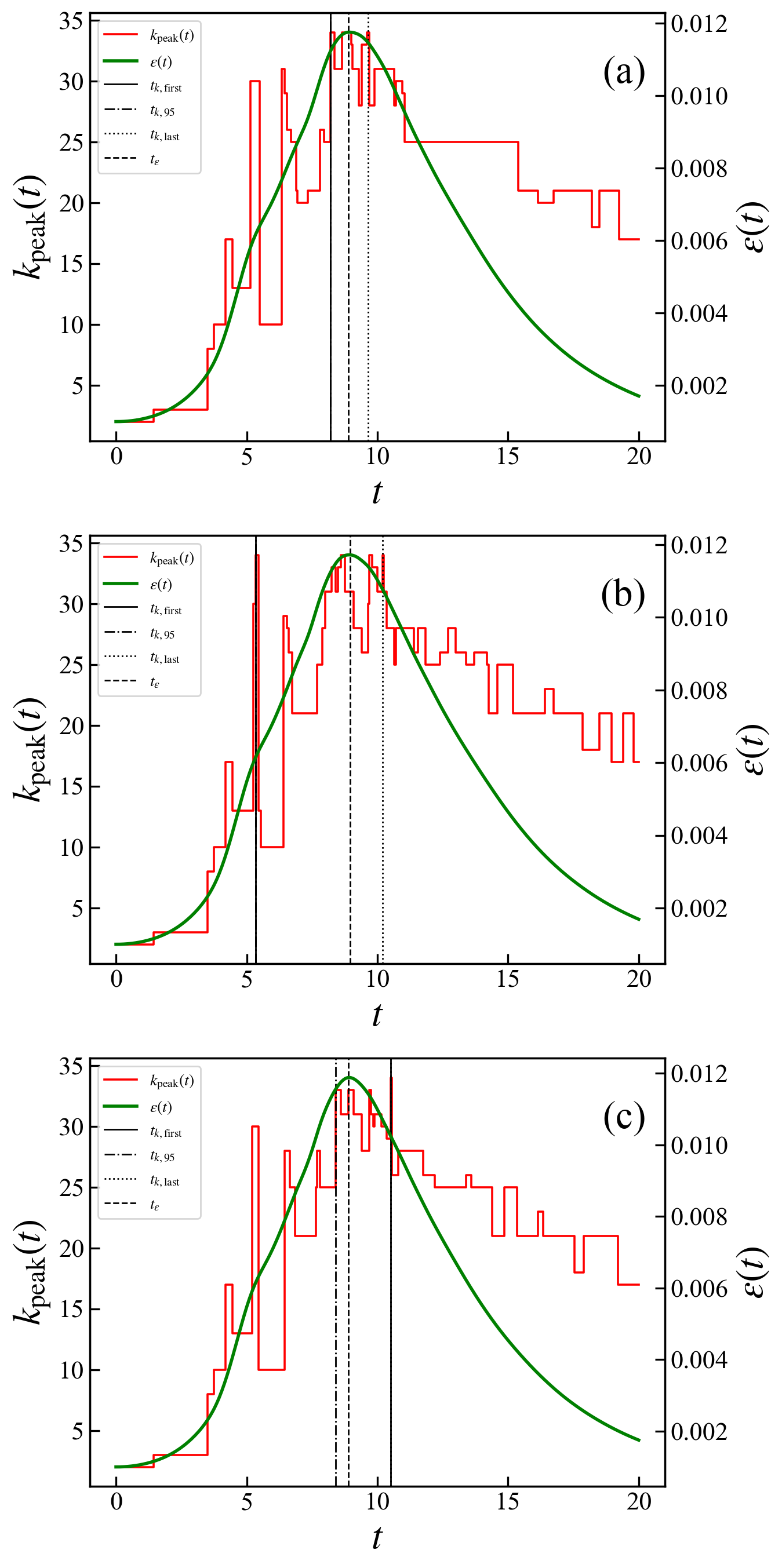}
  \caption{Example time series of the spectral peak wavenumber $k_{\mathrm{peak}}(t)$ (red) and viscous dissipation rate $\varepsilon(t)$ (green), illustrating typical, leading, and lagging realizations.
  Vertical lines indicate the timing markers $t_{k,\mathrm{first}}$, $t_{k,95}$, $t_{k,\mathrm{last}}$, and $t_\varepsilon$.}
  \label{fig:fig1}
\end{figure*}

Figure~\ref{fig:fig1} shows representative time series of $k_{\mathrm{peak}}(t)$ and $\varepsilon(t)$, together with the timing markers $t_{k,\mathrm{first}}$, $t_{k,95}$, $t_{k,\mathrm{last}}$, and $t_\varepsilon$.
The panels illustrate that (i) $k_{\mathrm{peak}}(t)$ typically increases as the cascade populates higher wavenumbers, (ii) the maximum $K_{\max}$ is often attained over a finite plateau, and (iii) the ordering between $t_k$ and $t_\varepsilon$ depends on which operational definition of $t_k$ is adopted. In particular, $t_{k,\mathrm{first}}$ and $t_{k,95}$ are designed to emphasize the onset of the plateau and thus act as precursors, whereas $t_{k,\mathrm{last}}$ can occur after $t_\varepsilon$ and is therefore not intended as a leading indicator.

\subsection{Distribution of $K_{\max}$ and conditional lag probability}
\begin{figure*}[t]
  \centering
  \includegraphics[width=\linewidth]{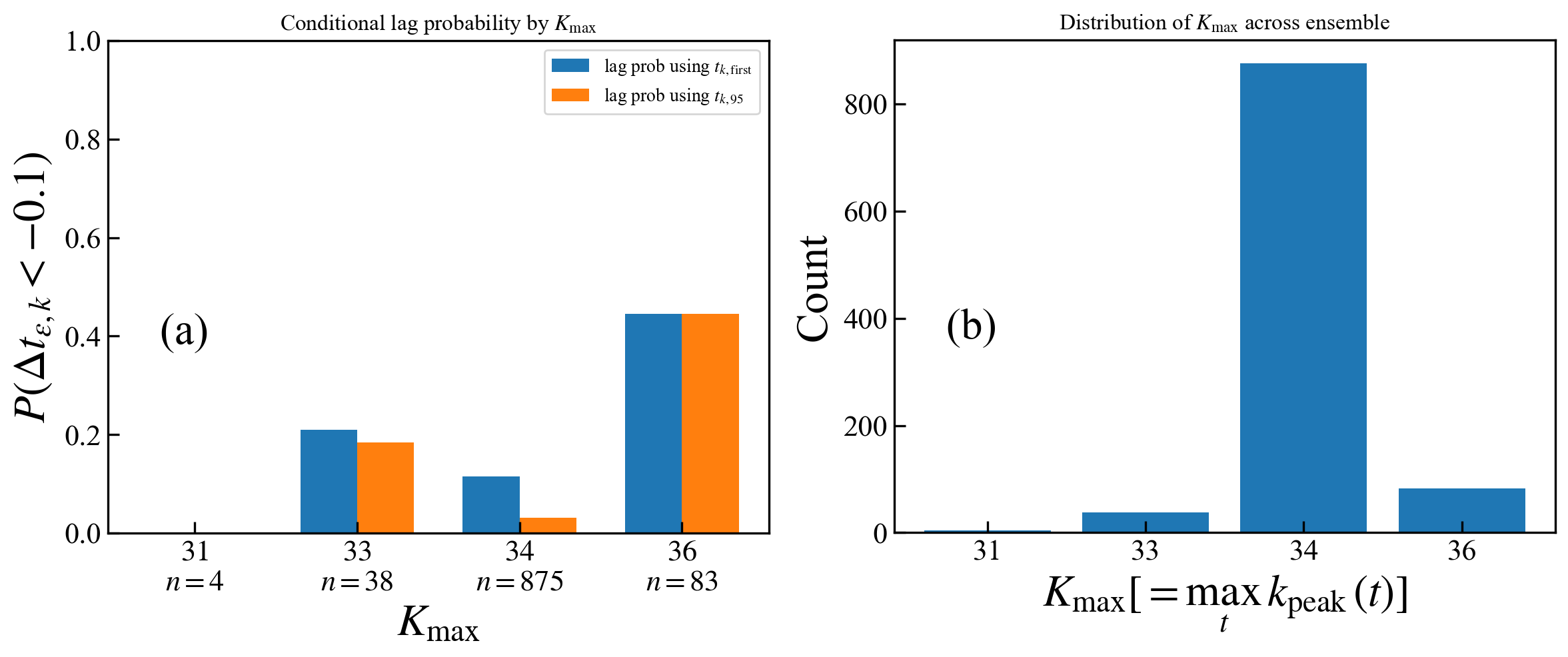}
  \caption{(a) Conditional lag probability $\Pr(\Delta t_{\varepsilon,k}<-0.1)$ as a function of $K_{\max}$, comparing $t_{k,\mathrm{first}}$ and $t_{k,95}$.
  (b) Histogram of $K_{\max}$ across the $N_s=1000$ ensemble.}
  \label{fig:fig2}
\end{figure*}

Figure~\ref{fig:fig2} summarizes the ensemble variability of $K_{\max}$ and the conditional probability of ``lag events'' $\Delta t_{\varepsilon,k}<-0.1$ given $K_{\max}$.
For the present ensemble ($N_s=1000$), $K_{\max}$ takes a small set of discrete values (31, 33, 34, 36), with the majority at $K_{\max}=34$.
Quantitatively, the observed $K_{\max}$ counts are as follows:
\begin{center}
\begin{tabular}{ccc}
\toprule
$K_{\max}$ & count & fraction \\
\midrule
31 & 4   & 0.004 \\
33 & 38  & 0.038 \\
34 & 875 & 0.875 \\
36 & 83  & 0.083 \\
\bottomrule
\end{tabular}
\end{center}
Two trends are apparent in Fig.~\ref{fig:fig2}(a):
(i) most realizations fall into $K_{\max}=34$, where the lag probability is modest and can be strongly reduced by adopting the fractional definition $t_{k,95}$ instead of $t_{k,\mathrm{first}}$; and
(ii) realizations with $K_{\max}=36$ are substantially more likely to exhibit lag, suggesting that the large-$K_{\max}$ branch corresponds to a distinct dynamical pathway (or to a different discretization regime of the spectral-peak trajectory; see Appendix \ref{app:specinterKmax}).

\subsection{Ensemble distributions of precursor lead time and plateau duration}
\begin{figure*}[t]
  \centering
  \includegraphics[width=\linewidth]{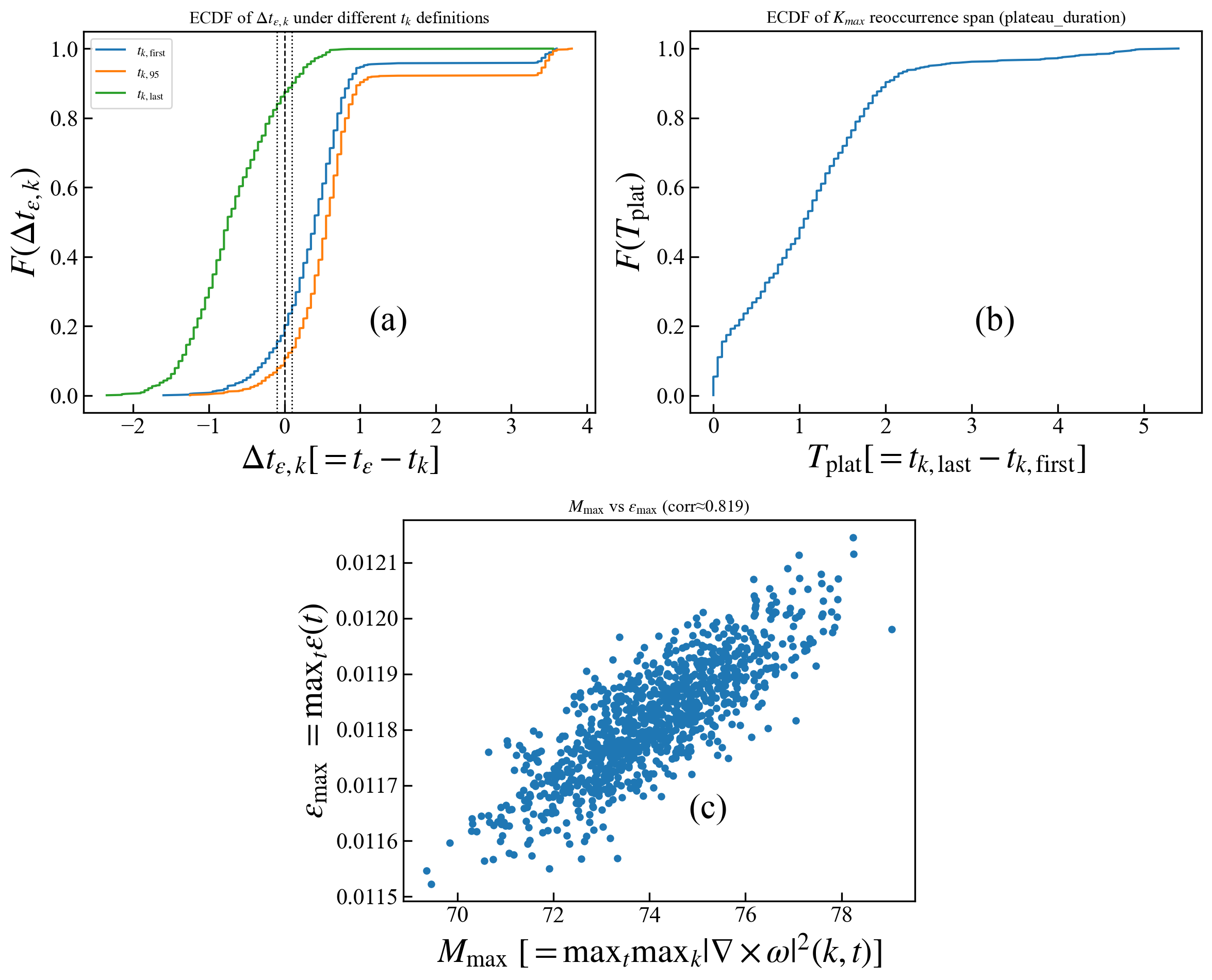}
  \caption{(a) ECDF of $\Delta t_{\varepsilon,k}=t_\varepsilon-t_k$ for three $t_k$ definitions.
  (b) ECDF of the $K_{\max}$ plateau duration $T_{\mathrm{plat}}=t_{k,\mathrm{last}}-t_{k,\mathrm{first}}$.
  (c) Scatter plot of $M_{\max}$ versus $\varepsilon_{\max}$ showing a strong correlation across the ensemble.}
  \label{fig:fig3}
\end{figure*}

Figure~\ref{fig:fig3}(a) shows the empirical cumulative distribution functions (ECDFs) of $\Delta t_{\varepsilon,k}$ for three $t_k$ definitions.
The distributions for $t_{k,\mathrm{first}}$ and $t_{k,95}$ are centered on positive values (typical lead), but they exhibit a negative tail corresponding to occasional lagging cases.
By contrast, the distribution for $t_{k,\mathrm{last}}$ is shifted to negative values, reflecting that the maximum-wavenumber plateau can persist beyond the dissipation peak.
Figure~\ref{fig:fig3}(b) shows the ECDF of the plateau duration $T_{\mathrm{plat}}$, which has broad support up to $\approx 5.5$ (nondimensional time units), indicating that $k_{\mathrm{peak}}(t)=K_{\max}$ often persists over an extended interval rather than at an isolated instant.
\HLS{Runs with exceptionally long plateaus should not be interpreted as remaining frozen at a single physical length scale. Because $k_{\mathrm{peak}}(t)$ is a shell-valued diagnostic, a long plateau means that the dominant shell of the curl-of-vorticity spectrum remains the same while the spectral amplitude and the detailed distribution within and around that shell continue to evolve. Physically, such cases correspond to a prolonged residence of the curvature-weighted spectral peak in the $K_{\max}$ shell, rather than to a stationary turbulent state.}
Finally, Fig.~\ref{fig:fig3}(c) shows that $M_{\max}$ and $\varepsilon_{\max}$ are strongly correlated across realizations (Pearson correlation $\approx 0.819$), implying that runs with stronger high-curvature activity also tend to realize stronger dissipation bursts.

\subsection{POT modeling of extreme lags and extreme $M_{\max}$}
\begin{figure*}[t]
  \centering
  \includegraphics[width=\linewidth]{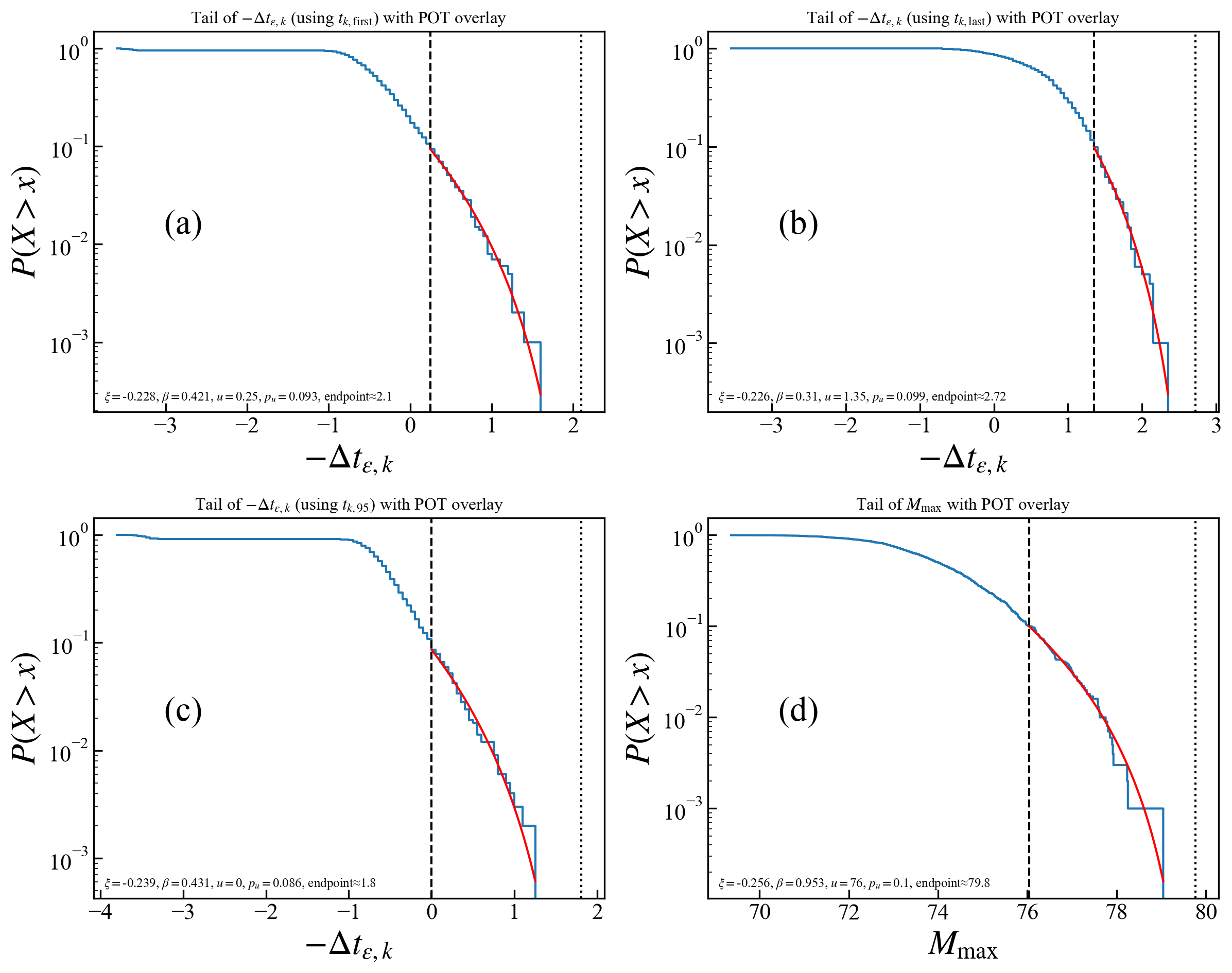}
  \caption{POT EVT analysis of upper tails.
  (a)--(c) Tail of the lag magnitude $X=-\Delta t_{\varepsilon,k}$ for different $t_k$ definitions, with GPD fit overlay.
  (d) Tail of $M_{\max}$ with GPD fit overlay.
  Dashed vertical lines indicate the threshold $u$ and dotted lines indicate the implied finite endpoint when $\hat\xi<0$.}
  \label{fig:fig4}
\end{figure*}

We now quantify the most extreme lag events using POT EVT.
For each timing definition, we consider the transformed variable $X=-\Delta t_{\varepsilon,k}$ and focus on its upper tail.
Figure~\ref{fig:fig4} shows the empirical survival function $\Pr(X>x)$ and the fitted GPD tail overlay.
\HLT{In all cases the fitted shape parameter is negative ($\hat\xi<0$), which is consistent with an effective bounded upper tail for the finite, discretely sampled reference ensemble. The corresponding endpoint estimates should be interpreted as protocol-dependent worst-case estimates at this resolution, not as strict bounds for the continuum dynamics.} We also apply POT to $M_{\max}$ and likewise obtain $\hat\xi<0$.
For transparency and reuse, Table~\ref{tab:potfits} summarizes the fitted POT parameters using a common quantile threshold $q=0.9$.
\begin{table*}[t]
\caption{The fitted POT parameters using a common quantile threshold $q = 0.9$}
\label{tab:potfits}
\begin{ruledtabular}
\begin{tabular}{lrrrrrrr}
Variable & $n$ & $q$ & $u$ & $p_u$ & $\hat\xi$ & $\hat\beta$ & $\hat x_{\mathrm{end}}$ \\
\midrule
$-\Delta t_{\varepsilon,k}^{\mathrm{(first)}}$ & 1000 & 0.90 & 0.25 & 0.093 & -0.228 & 0.421 & 2.10 \\
$-\Delta t_{\varepsilon,k}^{(s=0.95)}$        & 1000 & 0.90 & 0.00 & 0.086 & -0.239 & 0.431 & 1.80 \\
$-\Delta t_{\varepsilon,k}^{\mathrm{(last)}}$ & 1000 & 0.90 & 1.35 & 0.099 & -0.226 & 0.310 & 2.72 \\
$M_{\max}$                                     & 1000 & 0.90 & 76.0 & 0.100 & -0.256 & 0.953 & 79.8 \\
\end{tabular}
\end{ruledtabular}
\end{table*}
To assess threshold sensitivity and parameter uncertainty, we additionally performed fixed-threshold bootstrap resampling for each POT fit.
Table~\ref{tab:bootstrap_ci} reports 95\% confidence intervals for $(\xi,\beta,x_{\mathrm{end}})$, and Figs.~\ref{fig:fig7}--\ref{fig:fig9} provide complementary diagnostics (exceedance--threshold tradeoff, bootstrap histograms, and mean residual life plots). \HLS{A threshold-stability scan over empirical quantiles $q\in[0.85,0.95]$ is reported in Appendix~\ref{app:pot_diagnostics} and Fig.~\ref{fig:fig11}.}

\subsection{Time-resolved relationship between $M(t)$ and $\varepsilon(t)$}
\begin{figure*}[t]
  \centering
  \includegraphics[width=\linewidth]{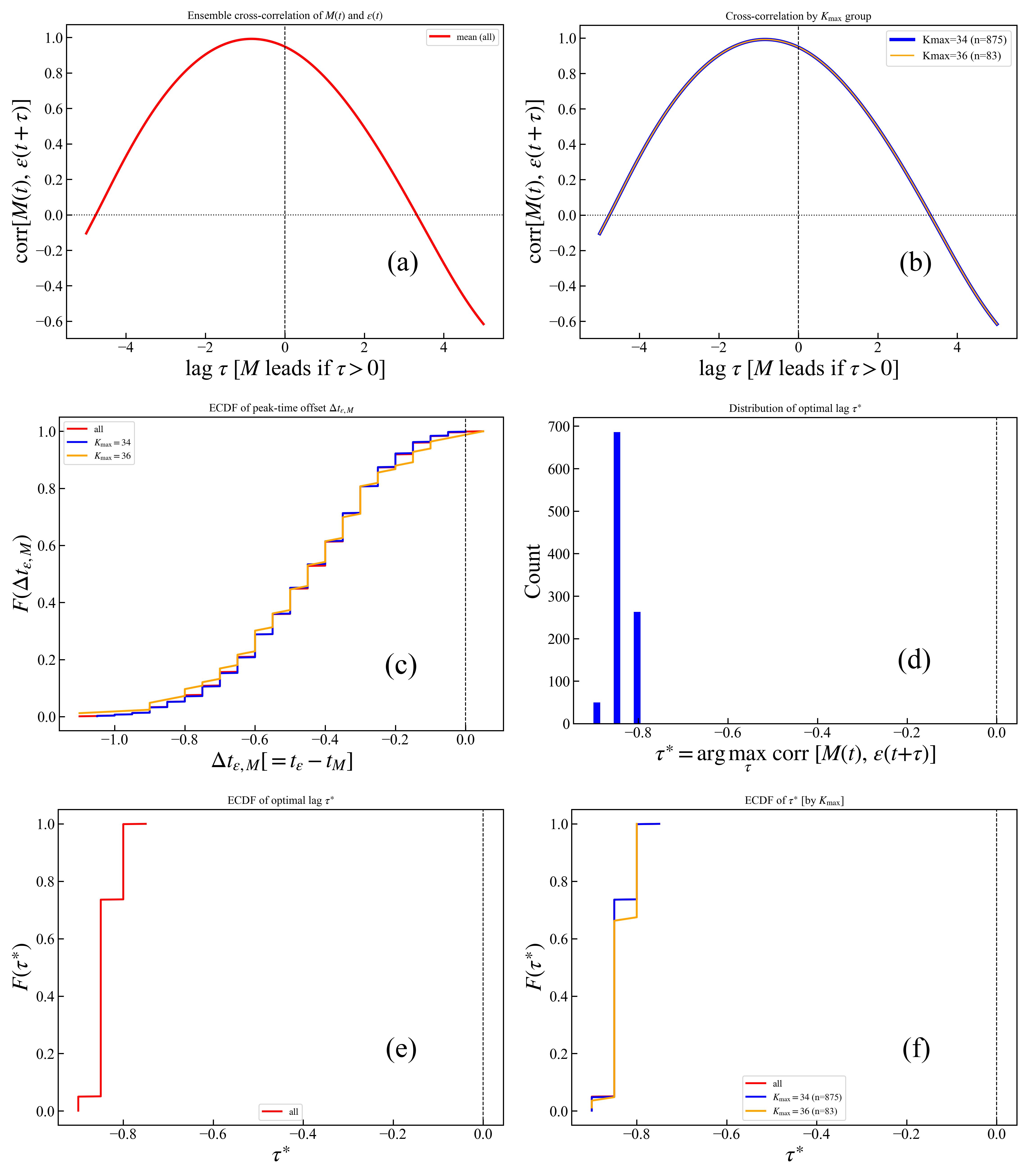}
  \caption{Time-resolved relationship between the curl-of-vorticity amplitude $M(t)$ and dissipation $\varepsilon(t)$.
  (a) Ensemble-mean cross-correlation $\mathrm{corr}[M(t),\varepsilon(t+\tau)]$.
  (b) Cross-correlation by $K_{\max}$ group (dominant groups $K_{\max}=34$ and $36$ shown).
  (c) ECDF of $\Delta t_{\varepsilon,M}=t_\varepsilon-t_M$ (negative indicates $M$ peaks after $\varepsilon$).
  (d)--(f) Distribution and ECDFs of the optimal lag $\tau^\ast=\arg\max_\tau \,\mathrm{corr}[M(t),\varepsilon(t+\tau)]$.}
  \label{fig:fig5}
\end{figure*}

Figure~\ref{fig:fig5} reports an auxiliary analysis based on time series written at the spectral sampling cadence.
We compute the ensemble cross-correlation $\mathrm{corr}[M(t),\varepsilon(t+\tau)]$ and identify the lag $\tau^\ast$ that maximizes this correlation for each realization. Across all realizations, $\tau^\ast$ is narrowly distributed around $\approx -0.85$, implying that $\varepsilon(t)$ tends to lead $M(t)$ by about $0.85$. \HLF{This ordering concerns the \emph{amplitude} $M(t)$, whereas the precursor studied in Figs.~\ref{fig:fig1}--\ref{fig:fig4} concerns the \emph{wavenumber location} $k_{\mathrm{peak}}(t)$.}
The two diagnostics therefore capture different aspects of the approach to the dissipation peak: scale localization (shift of $k_{\mathrm{peak}}$) versus intensity growth (rise of $M$).

\subsection{Unresolved low-viscosity case (spike-flagged)}
To clarify the limits of the precursor analysis, we also performed simulations at a lower viscosity $\nu=2.5\times 10^{-4}$.
At $N=256^3$, all realizations are flagged by spike inspection: the peak wavenumber locks at the de-aliased isotropic cutoff ($K_{\max}=86$ for this configuration), indicating insufficient resolution for reliable statistics of the curl-of-vorticity spectrum.
A smaller set of $N=512^3$ runs (not analyzed statistically here) exhibits the same failure mode, suggesting that substantially higher resolution would be required for this viscosity.
We therefore treat the low-viscosity dataset as a diagnostic cautionary example rather than as a target for EVT modeling.

\section{Discussion}\label{sec:discussion}
The DNS dynamics and the spectral post-processing are deterministic once the initial condition is fixed.
Nevertheless, for both experiments and computations, one rarely has access to the exact ``true'' initial condition: small perturbations (measurement noise, unmodeled background fluctuations, discretization and calibration uncertainty) imply that the precursor outcomes should be treated as \emph{random variables} sampled from an ensemble.
From the perspective of statistical physics, the present study therefore asks a concrete question:
\emph{given a well-defined perturbation ensemble around a canonical turbulent flow (TGV), what can extreme-value statistics say about the reliability and the worst-case behavior of a spectral precursor?}

\subsection{From a deterministic precursor to probabilistic reliability}
For a single realization, the ordering $t_k<t_\eps$ is a statement about a particular trajectory.
Across the ensemble, the relevant objects are the distributions of
(i) the scale indicator $k_{\mathrm{peak}}(t)$ and its hitting times $t_{k,\mathrm{first}}$, $t_{k,95}$, and $t_{k,\mathrm{last}}$,
(ii) the lag variables $\Delta t_{\eps,k}=t_\eps-t_k$ under these definitions,
and (iii) run-wise intensity measures such as $M_{\max}$ and $\eps_{\max}$.
Figure~\ref{fig:fig1} illustrates why these objects are not redundant:
the \emph{scale} at which the curl-of-vorticity spectrum concentrates (tracked by $k_{\mathrm{peak}}$) can evolve in a step-like fashion and produce multiple plausible ``event times'' even when $\eps(t)$ remains smooth.
In this setting, ``precursor reliability'' can be expressed in an operational form, e.g. as
$\mathbb{P}(\Delta t_{\eps,k}<-\delta)$ for a tolerance $\delta$ representing the minimal lead time that can be resolved or exploited.
The conditional probabilities in Fig.~\ref{fig:fig2}(a) show that this reliability is not a single number: it is strongly \emph{state dependent}.

\subsection{A discrete state variable: why conditioning on $K_{\max}$ is informative}
A salient empirical feature is that $\Kmax=\max_t k_{\mathrm{peak}}(t)$ takes only a small number of discrete values at the present resolution.
For the $N_s=1000$ reference ensemble, the distribution is dominated by $\Kmax=34$ (875 runs, $87.5\%$), with secondary branches at $\Kmax=36$ (83 runs, $8.3\%$), $\Kmax=33$ (38 runs, $3.8\%$), and a very small remainder at $\Kmax=31$ (4 runs); see Fig.~\ref{fig:fig2}(b).
This discreteness is not a drawback; it is a \emph{diagnostic}: it partitions the ensemble into distinct regimes with markedly different failure rates.
In particular, Fig.~\ref{fig:fig2}(a) shows that the lag probability $\mathbb{P}(\Delta t_{\eps,k}< -0.1)$ is smallest in the dominant $\Kmax=34$ regime, while it is substantially elevated in the $\Kmax=36$ branch and is also non-negligible in the $\Kmax=33$ branch.
This structure is difficult to infer from unconditional summary statistics alone.
From a modeling viewpoint, the results therefore favor a \emph{mixture} (or stratified) interpretation: the ensemble is better described as a superposition of sub-populations indexed by a discrete spectral state variable $\Kmax$.
A practical implication follows. If a measurement protocol (or online diagnostic) can determine whether a realization is evolving toward a higher-$\Kmax$ branch, then the inferred ``risk'' of precursor failure can be updated in a regime-aware manner.
In this sense, $\Kmax$ acts as a compact ``state label'' that carries predictive information beyond the instantaneous lead time itself.
\HLT{The discrete $K_{\max}$ branches observed here should be interpreted as conditional on the Taylor--Green geometry, the shell-averaged spectral diagnostic, the adopted resolution, and the perturbation ensemble. They are not claimed to be universal turbulence states. The result instead shows that, even within a fixed canonical flow, precursor reliability is not described by a single unconditional probability but is organized by discrete spectral-state labels.}

\subsection{Robust timing definitions and the role of the $k_{\mathrm{peak}}$ plateau}
The ECDFs in Fig.~\ref{fig:fig3}(a) confirm that the choice of $t_k$ definition matters.
The last-hitting time $t_{k,\mathrm{last}}$ produces systematically smaller $\Delta t_{\eps,k}$ (i.e. later inferred precursor times), consistent with the plateau interpretation:
once $k_{\mathrm{peak}}(t)$ reaches $\Kmax$, it can remain there for a finite span, and the \emph{duration} of this plateau controls how far $t_{k,\mathrm{last}}$ can be delayed relative to $t_{k,\mathrm{first}}$.
Figure~\ref{fig:fig3}(b) quantifies this reoccurrence span and shows that plateau durations are broadly distributed and can extend over several turnover-time fractions.
In this light, $t_{k,95}$ can be viewed as a compromise between early sensitivity and late-time ambiguity.
Operationally, it corresponds to the first time that the scale indicator enters a high-$k$ band close to $\Kmax$, suppressing spurious early steps while remaining insensitive to long plateaus.
The reduction of the conditional lag probability in the dominant $\Kmax=34$ regime (Fig.~\ref{fig:fig2}(a)) provides an empirical validation of this robustness criterion.

\subsection{Extreme-event viewpoint: bounded tails, uncertainty, and threshold diagnostics}
A main contribution of extreme-value theory (EVT) in the present context is that it targets the rare events that most directly affect \emph{risk}:
large negative $\Delta t_{\eps,k}$ (i.e.\ cases where the precursor time trails the dissipation peak by an amount that would compromise predictability).
We therefore modeled the tail of $X=-\Delta t_{\eps,k}$ using the POT framework (Fig.~\ref{fig:fig4}).
With $N_s=1000$, \HLT{the inference of effective bounded-tail behavior for the present finite-resolution ensemble} is substantially more robust than in the smaller ensemble:
(i) the POT fits yield $\hat{\xi}<0$ for all four variables considered, and
(ii) bootstrap confidence intervals for $\xi$ are strictly negative, with a bootstrap frequency of $\hat{\xi}\ge 0$ below 0.2 $\%$ for any variable (Table~\ref{tab:bootstrap_ci}).
\HLT{This supports the interpretation that, under the present measurement protocol and within the resolved reference regime, the right tails of $X=-\Delta t_{\eps,k}$ and $M_{\max}$ are effectively Weibull-type, without implying a continuum-level physical endpoint.}
At the same time, EVT diagnostics clarify an important trade-off.
Increasing the threshold $u$ reduces the number of exceedances $n_u$ and thus increases parametric uncertainty; Fig.~\ref{fig:fig7} makes this trade-off explicit for all four variables. \HLS{The additional threshold scan in Fig.~\ref{fig:fig11} shows that $\hat\xi$ remains negative over $q\in[0.85,0.95]$ for all variables, while the parameter estimates become noisier near the highest thresholds.}

The mean residual life (MRL) curves (Fig.~\ref{fig:fig9}) are broadly consistent with a decreasing, approximately linear trend at intermediate thresholds (as expected for $\xi<0$), but become progressively noisier as $u$ increases and $n_u$ shrinks.
These observations motivate a ``goldilocks'' threshold selection: high enough for the GPD approximation to be credible, but not so high that the fit becomes dominated by a handful of points.
Bootstrap histograms (Figs.~\ref{fig:fig7}--\ref{fig:fig8}) highlight that endpoint estimates can be sensitive to finite-sample variability, especially when $\hat{\xi}$ is close to zero (even if negative). \HLT{Endpoint estimates require particular caution because $x_{\mathrm{end}}=u-\beta/\xi$ is a nonlinear function of the fitted parameters and is sensitive when $\hat\xi$ is only mildly negative. In the present work, $x_{\mathrm{end}}$ is used as a diagnostic estimate of the largest plausible lag or amplitude under the adopted finite-resolution protocol, not as a sharply determined physical upper bound.}
For the present dataset, the fitted endpoints and their confidence intervals should therefore be interpreted as \emph{protocol- and sample-dependent worst-case bounds} rather than as strict continuum invariants.
The key point is not that the endpoint is known with high precision, but that bounded-tail behavior is consistently supported and that the implied worst-case lag magnitudes remain $\mathcal{O}(1)$ in the nondimensional time units used here.

\subsection{Physical meaning of curl-of-vorticity spectral intensity: coupling to dissipation and phase relations}
Beyond timing, the ensemble reveals a tight relationship between curl-of-vorticity activity and dissipation intensity:
Fig.~\ref{fig:fig3}(c) shows a strong positive correlation between $M_{\max}$ and $\eps_{\max}$ (Pearson $r \approx 0.819$ for $N_s=1000$).
Here $M(t)$ (and hence $M_{\max}$) is defined as the maximum over wavenumber of the curl-of-vorticity spectrum, i.e. $M(t)= \max_k \mathcal{C}(k,t)$, so it preferentially weights the smallest dynamically active scales captured by the spectrum.
The observed correlation therefore supports the interpretation that realizations with stronger high-curvature spectral activity also tend to realize stronger dissipation bursts, reinforcing the physical relevance of curl-of-vorticity diagnostics.
\HLT{The $M_{\max}$--$\varepsilon_{\max}$ correlation should not be read as an independent proof of predictive skill. It shows that the amplitude of the curl-of-vorticity spectrum co-varies with dissipation intensity, which supports the physical relevance of the diagnostic. The timing reliability of the precursor is instead assessed through the lead--lag variable $\Delta t_{\varepsilon,k}$, its conditional dependence on $K_{\max}$, and the POT analysis of rare lagging events.}

At the same time, the time-resolved cross-correlation analysis (Fig.~\ref{fig:fig5}) clarifies that ``physical relevance'' does not automatically imply ``causal precedence'' in time.
Both the peak-time offset $\Delta t_{\eps,M}=t_\eps-t_M$ and the optimal-lag statistics $\tau^\ast=\arg\max_\tau \mathrm{corr}[M(t),\eps(t+\tau)]$ indicate that $M(t)$ typically peaks \emph{after} $\eps(t)$ by a fraction of a unit time.
This is consistent with a picture in which the dissipation peak marks the culmination of energy removal, while the most intense higher-derivative spectral curvature measure continues to sharpen slightly later as the cascade reorganizes small-scale structures.
\HLF{This behavior can coexist with the precursor property of $k_{\mathrm{peak}}(t)$:}
the \emph{dominant scale} of curl-of-vorticity activity can reach near-saturation (affecting $t_k$) before the \emph{amplitude} $M(t)$ attains its maximum.
\HLT{The present analysis is empirical in the following sense. We do not derive a closed dynamical equation showing that migration of $k_{\mathrm{peak}}(t)$ causes the dissipation maximum. A plausible interpretation is that both $k_{\mathrm{peak}}(t)$ and $\varepsilon(t)$ respond to a common underlying process, such as vortex stretching, enstrophy production, and the transfer of activity toward smaller scales. A related physical picture is that the $k^4E(k)$ weighting makes $\mathcal{C}(k,t)$ sensitive to the migration of activity toward the smallest dynamically active scales, while dissipation responds to the $k^2E(k)$-weighted energy removal. This provides a natural reason why the two diagnostics are related, but it does not establish that the migration of $k_{\mathrm{peak}}(t)$ causes the dissipation maximum. The contribution of this work is therefore a statistical quantification of the reliability and failure modes of the observed timing relation, rather than a causal derivation of the precursor mechanism.}

\subsection{Limitations, resolution requirements, and outlook}
Finally, the low-viscosity case highlights a key methodological point: EVT cannot rescue under-resolved diagnostics.
Spike inspection is essential when analyzing high-$k$-sensitive indicators; otherwise, cutoff-proximate locking can masquerade as extreme precursor behavior. In the present work, this inspection justifies restricting EVT-based conclusions to the resolved reference case.
For completeness, we also report the time evolution of the classical resolution indicator $k_{\max}\eta$ (Appendix~\ref{app:kmaxeta}), which remains safely above unity at Nyquist for the reference dataset.

\HLT{All probabilities reported in this paper are conditional on the perturbation protocol $w=0.05$ and $k\le 3$. Changing the perturbation amplitude or allowing higher-wavenumber perturbations may alter the relative weights of the $K_{\max}$ branches and the corresponding lag probabilities. A systematic sweep over $w$ and perturbation bandwidth would require additional large ensembles and is left for future work. The present study should therefore be viewed as a statistical reliability analysis for a well-defined perturbation ensemble, rather than as a perturbation-model-independent statement.}

Several extensions are natural. On the physics side, it is important to test whether the mixture structure in $\Kmax$ persists at higher resolution and across different perturbation amplitudes and initial-condition families.
On the statistics side, the POT fits and their diagnostics (Appendix~\ref{app:pot_diagnostics}) can be extended to systematic threshold-selection criteria and to joint/conditional tail models (e.g. conditioning on $\Kmax$ or coupling $M_{\max}$ and $\eps_{\max}$ in a multivariate EVT framework).
More broadly, the methodology here provides a template for translating deterministic ``precursor'' observations into quantitative uncertainty bounds, which is a prerequisite for assessing observability and practical predictability in laboratory measurements and numerical models.

\section{Conclusions}
\label{sec:conclusions}

We investigated the statistical robustness of a turbulence precursor based on the \emph{curl-of-vorticity} spectrum in perturbed Taylor--Green vortex (TGV) turbulence.
Although the underlying pseudospectral DNS is deterministic, the precursor timing becomes practically uncertain once one accounts for small perturbations to the initial condition (representing experimental uncertainty, numerical uncertainty, or uncontrolled fluctuations).
To quantify this uncertainty, we performed an ensemble of $N_s=1000$ perturbed TGV realizations at $N=256^3$ and $\nu=10^{-3}$ and analyzed ensemble-level fluctuations using peaks-over-threshold (POT) extreme-value statistics.

\paragraph*{What we learned.}
Our main findings can be summarized as follows:
\begin{enumerate}[leftmargin=2.2em]
\item \textbf{Ensemble variability of precursor timing is measurable and highly structured.}
The lead time $\dt_{\eps,k}=t_\eps-t_k$ is broadly positive for the resolved reference case, but a statistically significant negative tail exists (rare ``precursor failures'' with $t_k>t_\eps$).
Moreover, the failure probability is strongly conditioned on the discrete wavenumber maximum $\Kmax=\max_t k_{\mathrm{peak}}(t)$.
For $N_s=1000$, the ensemble is dominated by $\Kmax=34$ (875 runs), with secondary branches at $\Kmax=36$ (83 runs) and $\Kmax=33$ (38 runs) (Fig.~\ref{fig:fig2}).
\HLT{The conditional lag probability $\mathbb{P}(\dt_{\eps,k}<-0.1)$ is smallest in the $\Kmax=34$ regime and is substantially larger in the higher-$\Kmax$ branch. Within the present TGV ensemble, this shows that precursor reliability is state dependent rather than captured by a single unconditional probability.}

\item \textbf{A more robust $t_k$ definition can be justified and validated statistically.}
Defining $t_k$ via the fractional criterion $t_{k,s}$ (here $s = 0.95$) reduces the incidence of lagging events in the dominant $\Kmax=34$ group while largely preserving the typical leading behavior (Fig.~\ref{fig:fig2}).
This provides a concrete, data-driven route to \emph{redefining} the precursor time so as to mitigate ambiguity caused by the finite plateau of $k_{\mathrm{peak}}(t)$ at $\Kmax$.

\item \textbf{The severity of rare lagging events admits EVT-based upper bounds, with quantified uncertainty.}
\HLT{Applying POT to the lag magnitude $X=-\dt_{\eps,k}$ yields negative GPD shape parameters for all three $t_k$ definitions, consistent with effective bounded right tails for the finite, discretely sampled reference ensemble.}
Using a fixed $q=0.9$ threshold and bootstrap resampling, we find strictly negative 95\% confidence intervals for $\xi$ for all variables, and a bootstrap frequency of $\hat{\xi}\ge 0$ below 0.2 $\%$ in all cases (Table~\ref{tab:bootstrap_ci}).
Endpoint estimates are finite but have non-negligible uncertainty, reflecting the expected exceedance/variance trade-off at high thresholds (Appendix~\ref{app:pot_diagnostics}); nevertheless, the fitted endpoints remain $\mathcal{O}(1)$ for $X=-\dt_{\eps,k}$ in the present nondimensional units.

\item \HLT{\textbf{The curl-of-vorticity spectral intensity co-varies with dissipation bursts but does not by itself establish predictive skill.}}
Across realizations, the run-wise curl-of-vorticity spectral intensity
$M_{\max}=\max_t\max_k \mathcal{C}(k,t)$
\HLT{correlates strongly with the dissipation peak $\eps_{\max}$ (Pearson $r \approx 0.82$), indicating that high-curvature spectral activity co-varies with an energetically meaningful quantity, but the amplitude correlation alone does not demonstrate precursor skill.}
Time-resolved analysis further shows a reproducible phase offset between $M(t)$ and $\eps(t)$ (Fig.~\ref{fig:fig5}), with $M(t)$ typically peaking after $\eps(t)$, while the \emph{wavenumber localization} signal $k_{\mathrm{peak}}(t)$ remains a viable precursor in many realizations.

\item \textbf{Resolution screening is essential for preventing spurious EVT conclusions in under-resolved cases.}
For the lower-viscosity case $\nu=2.5\times 10^{-4}$, spike inspection flags cutoff-proximate peak locking at both $N=256^3$ and $N=512^3$, indicating that the spectra are not adequately resolved for reliable precursor/EVT inference.
This motivates a resolution-first workflow: only ensembles passing spike inspection (and classical criteria such as $k_{\max}\eta$) should be subjected to tail modeling.
\end{enumerate}

\paragraph*{Outlook.}
The present study opens several concrete directions.
First, extending the ensemble to lower viscosities will require systematically increasing resolution (and possibly refining dealiasing/cutoff choices) to avoid peak locking.
Second, the observed conditioning on $\Kmax$ motivates stratified (or hierarchical) EVT models that explicitly incorporate discrete spectral states.
\HLT{Third, the strong $M_{\max}$--$\eps_{\max}$ correlation and the reproducible phase relation between $M(t)$ and $\eps(t)$ suggest an empirical association between high-curvature activity and dissipation events. Establishing a causal mechanism will require joint real- and spectral-space diagnostics (e.g. structure geometry, enstrophy production, and local strain--vorticity alignment) and, potentially, multivariate or conditional EVT approaches.}
Finally, the methodology here provides a template for translating deterministic ``precursor'' observations into quantitative uncertainty bounds, which is a prerequisite for assessing observability and practical predictability in laboratory measurements and numerical models.

\begin{acknowledgments}
This study was supported by JSPS KAKENHI (Grant Number 22K14177) and JST PRESTO (Grant Number JPMJPR23O7).
\end{acknowledgments}
\appendix

\section{Resolution diagnostic based on $k_{\max}\eta$}
\label{app:kmaxeta}
\begin{figure*}[t]
  \centering
  \includegraphics[width=0.85\linewidth]{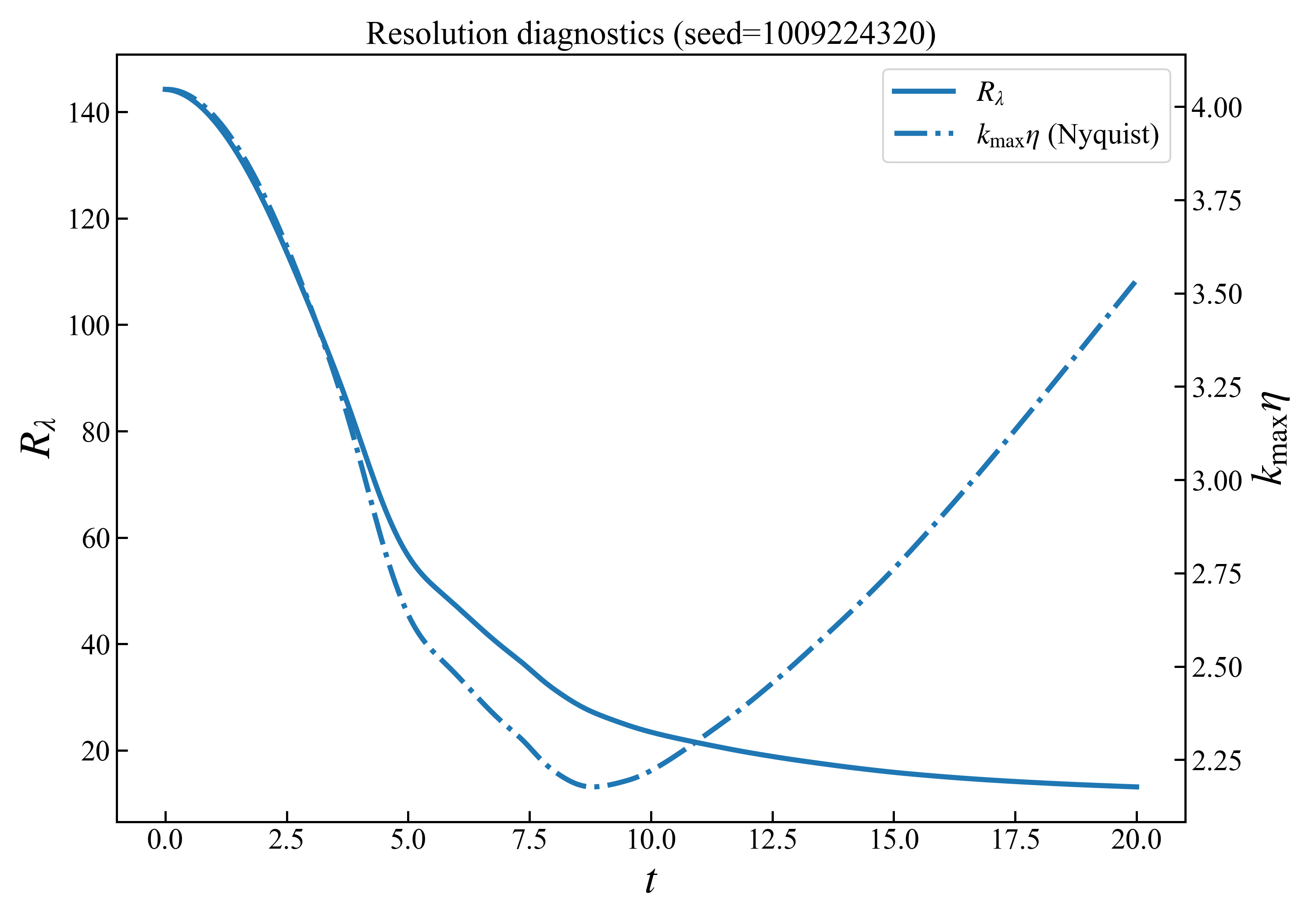}
  \caption{Representative time series (seed $=1009224320$) of the Taylor-microscale Reynolds number $R_{\lambda}(t)$ (solid line, left axis) and the resolution diagnostic $k_{\max}\eta(t)$ (dash-dotted line, right axis) for the perturbed TGV ensemble at $N=256^3$ and $\nu=10^{-3}$. The minimum of $k_{\max}\eta(t)$ occurs near the dissipation peak, where the smallest dissipative scales are expected, and subsequently increases as the flow decays.}
  \label{fig:fig6}
\end{figure*}

To complement the spike inspection, we report the temporal evolution of
the Taylor-microscale Reynolds number $R_{\lambda}(t)$ together with the
resolution indicator $k_{\max}\eta(t)$, where $k_{\max}=N/2$ denotes the Nyquist
wavenumber and $\eta(t)=(\nu^{3}/\varepsilon(t))^{1/4}$ is computed from the
spectral dissipation rate $\varepsilon(t)=2\nu\sum_{k}k^{2}E(k,t)$.
Figure~\ref{fig:fig6} shows a representative case from the ensemble
(seed $=1009224320$). As expected for decaying turbulence, $k_{\max}\eta(t)$ reaches
its minimum near the dissipation peak and then increases as the flow decays.
Across all $N_s=1000$ runs, $\min_t(k_{\max}\eta)$ is tightly distributed in
$[2.165,\,2.194]$ with median $2.180$ (IQR $[2.177,\,2.183]$), and the minimum
is attained at $t_\varepsilon$ in every run. Using the $2/3$ de-aliased cutoff
$k_{\rm cut}=N/3$, the corresponding minimum satisfies $\min_t(k_{\rm cut}\eta)\in[1.443,\,1.463]$
(median $1.453$), consistent with $k_{\rm cut}\eta=(2/3)k_{\max}\eta$.

\section{Spectral interpretation of the discrete $K_{\max}$ branches}\label{app:specinterKmax}
To clarify how the discrete values of $K_{\max}$ arise and how they relate to the regime dependence reported in Fig.~\ref{fig:fig2}, Fig.~\ref{fig:fig10} visualizes the time evolution of the spectra for three representative realizations with $K_{\max}=33$, $34$, and $36$.
Here $C(k,t)$ denotes the isotropic curl-of-vorticity spectrum (proportional to $k^4E(k,t)$ for incompressible flow), and $k_{\mathrm{peak}}(t)=\arg\max_k C(k,t)$ is shown by red markers.
The vertical dotted line indicates the run-wise maximum $K_{\max}=\max_t k_{\mathrm{peak}}(t)$.

Across all three cases, the red markers demonstrate that $k_{\mathrm{peak}}(t)$ intermittently approaches $K_{\max}$ and later departs from it, implying that the maximum-wavenumber state is visited repeatedly rather than at a single instant.
The superposed black curves in the zoomed panels show $C(k,t)$ evaluated at the extracted event times $t_{k,\mathrm{first}}$ and $t_{k,\mathrm{last}}$ (first and last hitting times of $K_{\max}$), together with $t_\varepsilon$ and $t_M$.
As expected from the operational definitions, the spectra at $t_{k,\mathrm{first}}$ and $t_{k,\mathrm{last}}$ attain their maxima at $k=K_{\max}$; the purpose of this overlay is to provide a direct visual sanity check of the extraction procedure and to prevent confusion between $K_{\max}$ and other peak-related quantities.

\HLF{$K_{\max}$ should not be conflated with the wavenumber at which the global amplitude maximum occurs.}
The maximum spectral amplitude is $M_{\max}=\max_t \max_k C(k,t)$ at time $t_M$, and its associated peak location is $k_{\mathrm{peak}}(t_M)$, which need not coincide with $K_{\max}$ (e.g., in the representative $K_{\max}=34$ case, $k_{\mathrm{peak}}(t_M)=31$).
Comparing the three realizations, the $K_{\max}=36$ case exhibits a noticeable bias of the curvature-weighted activity toward smaller scales: the peak location at the maximum-amplitude time is shifted to a higher wavenumber ($k_{\mathrm{peak}}(t_M)=34$) compared with the dominant $K_{\max}=34$ case.
This trend is consistent with the conditional statistics in Fig.~\ref{fig:fig2}, where the high-$K_{\max}$ branch shows a substantially elevated probability of lagging precursor times ($\Delta t_{\varepsilon,k}<0$), suggesting that realizations that access a higher-$K_{\max}$ pathway are more prone to delayed migration of $k_{\mathrm{peak}}(t)$ relative to the dissipation peak.

\HLS{To characterize the long-plateau cases more explicitly, Fig.~\ref{fig:fig12} compares $T_{\mathrm{plat}}$ with $K_{\max}$, the lead--lag variables, $M_{\max}$, and $\varepsilon_{\max}$. The upper $5\%$ threshold is $T_{\mathrm{plat}}\ge 2.5025$, giving 50 long-plateau cases. These cases occur mainly in the dominant $K_{\max}=34$ branch, with no long-plateau cases in the $K_{\max}=36$ branch. The correlations of $T_{\mathrm{plat}}$ with $M_{\max}$ and $\varepsilon_{\max}$ are weak, indicating that long plateaus are not primarily markers of unusually intense dissipation or unusually large curl-of-vorticity spectral amplitude. Instead, they quantify how long the peak shell remains at $K_{\max}$ after reaching it.}

\section{Pickands--Balkema--de Haan theorem and generalized Pareto distributions}
\label{app:gpd_excess}
The following appendix provides a self-contained discussion of the theoretical basis for the POT method used in Sec.~\ref{sec:evt}.

\subsection*{Conditional excess distributions.}
Let $X$ be a real-valued random variable with distribution function $F(x)=\Pr(X\le x)$ and survival function $\bar F(x)=1-F(x)$.
Let the (right) endpoint be
\begin{equation}
  x_F := \sup\{x\in\mathbb{R}:F(x)<1\}\in(-\infty,\infty].
\end{equation}
In the main text, we denote by $x_{\mathrm{end}}$ the finite upper endpoint implied by the fitted GPD under $\xi < 0$. In the asymptotic limit $u \rightarrow x_F$, this estimator approximates the theoretical right endpoint $x_F$.

Fix a threshold $u<x_F$ such that $\bar F(u)>0$.
The \emph{(conditional) excess distribution} above $u$ is defined for $y\ge 0$ (with $u+y<x_F$) by
\begin{equation}
  F_u(y) := \Pr(X-u\le y \mid X>u).
  \label{eq:Fu_def}
\end{equation}
By the definition of conditional probability,
\begin{align}
  F_u(y)
  &= \frac{\Pr(u<X\le u+y)}{\Pr(X>u)}
   = \frac{F(u+y)-F(u)}{\bar F(u)}.
  \label{eq:Fu_basic}
\end{align}
Equivalently, the conditional survival function of the exceedance is the \emph{tail ratio}
\begin{align}
  \bar F_u(y)
  &:= 1-F_u(y) \nonumber \\
  &~= \Pr(X-u>y\mid X>u) \nonumber \\
  &~= \frac{\Pr(X>u+y)}{\Pr(X>u)} \nonumber \\
  &~= \frac{\bar F(u+y)}{\bar F(u)}.
  \label{eq:tail_ratio}
\end{align}
The PBdH theorem characterizes when $\bar F_u(\cdot)$ has an approximately universal form for \emph{high} thresholds $u$.

\subsection*{Generalized Pareto distribution and POT}
The generalized Pareto distribution (GPD) with shape parameter $\xi\in\mathbb{R}$ and scale $\beta>0$ has distribution function
\begin{equation}
  H_{\xi, \beta}(y) =
  \begin{cases}
    1 - \left(1+\xi y/\beta\right)^{-1/\xi}, & \xi\ne 0,\\[4pt]
    1 - \exp\left(-y/\beta\right), & \xi=0,
  \end{cases}
  \label{eq:GPD_cdf}
\end{equation}
defined for $y\ge 0$ subject to $1+\xi y/\beta>0$ (so that for $\xi<0$ the support is $0\le y\le -\beta/\xi$).
A key property is \emph{threshold stability}: if $Y\sim\mathrm{GPD}(\xi,\beta)$ and $t>0$ is such that $\Pr(Y>t)>0$, then
\begin{equation}
  (Y-t \mid Y>t) \sim \mathrm{GPD}(\xi,\beta+\xi t),
  \label{eq:threshold_stability}
\end{equation}
i.e., the shape $\xi$ is invariant under further thresholding while the scale updates linearly~\cite{Coles2001,EmbrechtsKluppelbergMikosch1997}.
In POT modeling, one chooses a high threshold $u$ and models the excess $Y=X-u$ conditional on $X>u$ as GPD.
The mathematical justification is provided by the PBdH theorem~\cite{BalkemaDeHaan1974,Pickands1975}.

\subsection*{Pickands--Balkema--de Haan theorem}
There are several equivalent formulations; we state a standard version emphasizing uniform convergence of the excess distribution.

Assume throughout this section that $(X_i)_{i\ge 1}$ are independent and identically distributed (i.i.d.) copies of $X$,
and define the sample maximum $M_n=\max\{X_1,\dots,X_n\}$.
The distribution $F$ is said to belong to the maximum domain of attraction of the generalized extreme-value (GEV) distribution with index $\xi$
if there exist normalizing sequences $a_n>0$ and $b_n\in\mathbb{R}$ such that
\begin{equation}
  \Pr\!\left(\frac{M_n-b_n}{a_n}\le z\right)\ \longrightarrow\ G_\xi(z),\qquad n\to\infty,
  \label{eq:MDA_def}
\end{equation}
where the GEV cdf is
\begin{equation}
  G_\xi(z)=
  \begin{cases}
    \exp\!\left(-(1+\xi z)^{-1/\xi}\right), & \xi\ne 0,\\[4pt]
    \exp\!\left(-e^{-z}\right), & \xi=0,
  \end{cases}
  \label{eq:GEV_cdf}
\end{equation}
with support restricted to $1+\xi z>0$ when $\xi\ne 0$~\cite{deHaanFerreira2006,Resnick1987}.
Under this assumption, high-threshold exceedances admit a GPD approximation~\cite{BalkemaDeHaan1974,Pickands1975}.

Assume that $X$ has distribution function $F$ in the \emph{maximum domain of attraction} of a GEV distribution with extreme-value index $\xi$,
denoted $F\in\mathrm{MDA}(\mathrm{GEV}_\xi)$~\cite{deHaanFerreira2006,Resnick1987}.
Then there exists a positive function $\beta(u)$ such that, as $u\to x_F$,
\begin{equation}
  \sup_{0\le y < x_F-u}
  \left|
    F_u(y) - H_{\xi,\beta(u)}(y)
  \right|
  \longrightarrow 0,
  \label{eq:PBdH}
\end{equation}
where $H_{\xi,\beta}$ is the GPD cdf in Eq.~\eqref{eq:GPD_cdf}.
Conversely, if the excess distributions satisfy Eq.~\eqref{eq:PBdH} for some $\xi$ and some positive $\beta(u)$, then $F\in\mathrm{MDA}(\mathrm{GEV}_\xi)$; thus the GPD limit for exceedances and the GEV limit for maxima are two sides of the same asymptotic regularity~\cite{deHaanFerreira2006,Resnick1987}.
In words, for sufficiently high thresholds, the conditional excess distribution is well approximated by a GPD whose shape parameter $\xi$
matches the GEV extreme-value index, while the scale $\beta(u)$ may depend on the threshold.
The case $\xi>0$ corresponds to heavy-tailed (Fr\'echet-type) behavior; $\xi=0$ to Gumbel-type; and $\xi<0$ to finite-endpoint (Weibull-type) behavior~\cite{deHaanFerreira2006,Coles2001}.

\subsection*{Derivation sketch via tail quantiles and extended regular variation}
This section sketches how the GPD form emerges from the tail ratio in Eq.~\eqref{eq:tail_ratio} under the MDA assumption.
A convenient tool is the \emph{tail quantile function}
\begin{equation}
  U(t) := F^{\leftarrow}\!\left(1-\frac{1}{t}\right),\qquad t>1,
  \label{eq:tail_quantile}
\end{equation}
where $F^{\leftarrow}$ is the generalized inverse (so that the definition remains valid even if $F$ is not strictly invertible)~\cite{deHaanFerreira2006}.
By construction, $\Pr(X>U(t))=1/t$.

\subsection*{Extended regular variation}
A cornerstone equivalence in extreme-value theory is that
\begin{equation}
  F\in\mathrm{MDA}(\mathrm{GEV}_\xi)
  \quad \Longleftrightarrow \quad
  U \in \mathrm{ERV}_\xi,
  \label{eq:MDA_ERV_equiv}
\end{equation}
where $\mathrm{ERV}_\xi$ denotes \emph{extended regular variation} with index $\xi$~\cite{deHaanFerreira2006,Resnick1987,BinghamGoldieTeugels1987}.
Concretely, $U\in \mathrm{ERV}_\xi$ means that there exists a scaling function $a(t)>0$ such that for each fixed $x>0$,
\begin{equation}
  \frac{U(tx)-U(t)}{a(t)}
  \longrightarrow
  \begin{cases}
    \displaystyle \frac{x^\xi-1}{\xi}, & \xi\ne 0,\\[6pt]
    \log x, & \xi=0,
  \end{cases}
  \qquad (t\to\infty).
  \label{eq:ERV}
\end{equation}
For the purposes of this note, Eq.~\eqref{eq:ERV} can be taken as a standard characterization of the MDA condition~\cite{deHaanFerreira2006}.

\subsection*{From ERV to the GPD tail ratio}
Choose a high threshold of the form
\begin{equation}
  u = U(t),\qquad t\to\infty.
  \label{eq:u_Ut}
\end{equation}
Consider normalized exceedances $(X-u)/a(t)$ conditional on $X>u$.
Using Eq.~\eqref{eq:tail_ratio}, for $y\ge 0$ we have
\begin{equation}
  \Pr\!\left(\frac{X-u}{a(t)} > y \,\Big|\, X>u\right)
  =
  \frac{\Pr(X>u+a(t)y)}{\Pr(X>u)}.
  \label{eq:cond_tail_scaled}
\end{equation}

\paragraph*{Case $\xi\ne 0$.}
Fix $y\ge 0$ and set
\begin{equation}
  x = (1+\xi y)^{1/\xi},
  \qquad\text{equivalently}\qquad
  y=\frac{x^\xi-1}{\xi}.
  \label{eq:x_choice}
\end{equation}
By ERV in Eq.~\eqref{eq:ERV}, $U(tx)\approx U(t)+a(t)y$, i.e.,
\begin{equation}
  u+a(t)y \approx U(tx).
  \label{eq:U_approx}
\end{equation}
Using $\Pr(X>U(t))=1/t$ and $\Pr(X>U(tx))=1/(tx)$,
\begin{equation}
  \frac{\Pr(X>u+a(t)y)}{\Pr(X>u)}
  \approx
  \frac{\Pr(X>U(tx))}{\Pr(X>U(t))}
  =
  \frac{1/(tx)}{1/t}
  =
  \frac{1}{x}.
  \label{eq:ratio_to_1overx}
\end{equation}
Substituting Eq.~\eqref{eq:x_choice} yields
\begin{equation}
  \Pr\!\left(\frac{X-u}{a(t)} > y \,\Big|\, X>u\right)
  \longrightarrow
  (1+\xi y)^{-1/\xi},
  \qquad (t\to\infty).
  \label{eq:GPD_tail_limit_xi}
\end{equation}
Taking complements, the corresponding conditional cdf converges to
\begin{equation}
  \Pr\!\left(\frac{X-u}{a(t)} \le y \,\Big|\, X>u\right)
  \longrightarrow
  1-(1+\xi y)^{-1/\xi},
  \label{eq:GPD_cdf_limit_xi}
\end{equation}
which is the GPD with unit scale in Eq.~\eqref{eq:GPD_cdf}.

\paragraph*{Case $\xi=0$.}
ERV gives $(U(tx)-U(t))/a(t)\to \log x$. Choose $x=e^{y}$ so that $\log x = y$.
Proceeding as above yields
\begin{equation}
  \Pr\!\left(\frac{X-u}{a(t)} > y \,\Big|\, X>u\right)
  \longrightarrow
  e^{-y},
  \label{eq:GPD_tail_limit_0}
\end{equation}
i.e., the exponential limit corresponding to the $\xi=0$ GPD.

\subsection*{Returning to unnormalized excesses}
The limits above are for the \emph{scaled} exceedance $(X-u)/a(t)$ with $u=U(t)$.
Equivalently, defining $\beta(u):=a(t)$ when $u=U(t)$, we may rewrite the approximation in the original scale as
\begin{equation}
  \Pr(X-u \le y \mid X>u)
  \approx
  H_{\xi,\beta(u)}(y)
  =
  1-\left(1+\xi \frac{y}{\beta(u)}\right)^{-1/\xi},
  \label{eq:unscaled_GPD_approx}
\end{equation}
with the exponential form for $\xi=0$.
This is the content of Eq.~\eqref{eq:PBdH}, and it explains why the GPD emerges as the universal limit for high-threshold exceedances.

\subsection*{Worked example: Pareto}
For the Pareto (Type I) distribution with parameters $x_m>0$ and $\alpha>0$,
\begin{equation}
  F(x)=1-\left(\frac{x_m}{x}\right)^\alpha,\qquad x\ge x_m,
  \label{eq:Pareto_cdf}
\end{equation}
the survival function is $\bar F(x)=(x_m/x)^\alpha$.
The tail quantile is explicit:
\begin{equation}
  U(t)=F^{-1}\!\left(1-\frac{1}{t}\right)=x_m t^{1/\alpha}.
\end{equation}
Setting $\xi=1/\alpha$ and $a(t)=(x_m/\alpha)t^{1/\alpha}$, one finds
\begin{equation}
  \frac{U(tx)-U(t)}{a(t)} = \frac{x^\xi-1}{\xi}
\end{equation}
exactly (no limiting argument needed), hence $F\in\mathrm{MDA}(\mathrm{GEV}_\xi)$.

Moreover, the excess distribution above any threshold $u\ge x_m$ is \emph{exactly} GPD:
using Eq.~\eqref{eq:tail_ratio},
\begin{equation}
  \bar F_u(y) = \frac{\bar F(u+y)}{\bar F(u)}
  = \left(\frac{u}{u+y}\right)^\alpha
  = \left(1+\frac{y}{u}\right)^{-\alpha}.
\end{equation}
Comparing with the GPD survival function $(1+\xi y/\beta)^{-1/\xi}$ shows that
\begin{equation}
  X-u \mid X>u \ \sim\ \mathrm{GPD}\!\left(\xi=\frac{1}{\alpha},\, \beta=\frac{u}{\alpha}\right).
  \label{eq:Pareto_exact_GPD}
\end{equation}
Thus the PBdH approximation becomes an identity for Pareto tails.

\subsection*{Practical remarks for applications.} The PBdH theorem is asymptotic in the threshold: $u\to x_F$ (or $u\to\infty$ if $x_F=\infty$).
In finite samples, one typically chooses a high threshold $u=u_n$ such that the number of exceedances $k_n$ satisfies
$k_n\to\infty$ while $k_n/n\to 0$ as $n\to\infty$ (an \emph{intermediate} sequence), balancing bias and variance~\cite{Coles2001,deHaanFerreira2006}.
When observations are dependent (e.g., time series), exceedances may cluster; declustering or extremal-index corrections are commonly used in POT practice~\cite{Coles2001,EmbrechtsKluppelbergMikosch1997}.

In conclusion, starting from the conditional excess distribution in Eq.~\eqref{eq:Fu_def}, the tail-ratio identity in Eq.~\eqref{eq:tail_ratio}
reduces the problem to understanding $\bar F(u+y)/\bar F(u)$ for large $u$.
Under the MDA/ERV regularity condition, the scaled exceedance tail converges to the GPD form, yielding the PBdH theorem and providing the mathematical basis for POT modeling of extremes.

\section{POT threshold and uncertainty diagnostics}\label{app:pot_diagnostics}
\begin{figure*}[t]
  \centering
  \includegraphics[width=\linewidth]{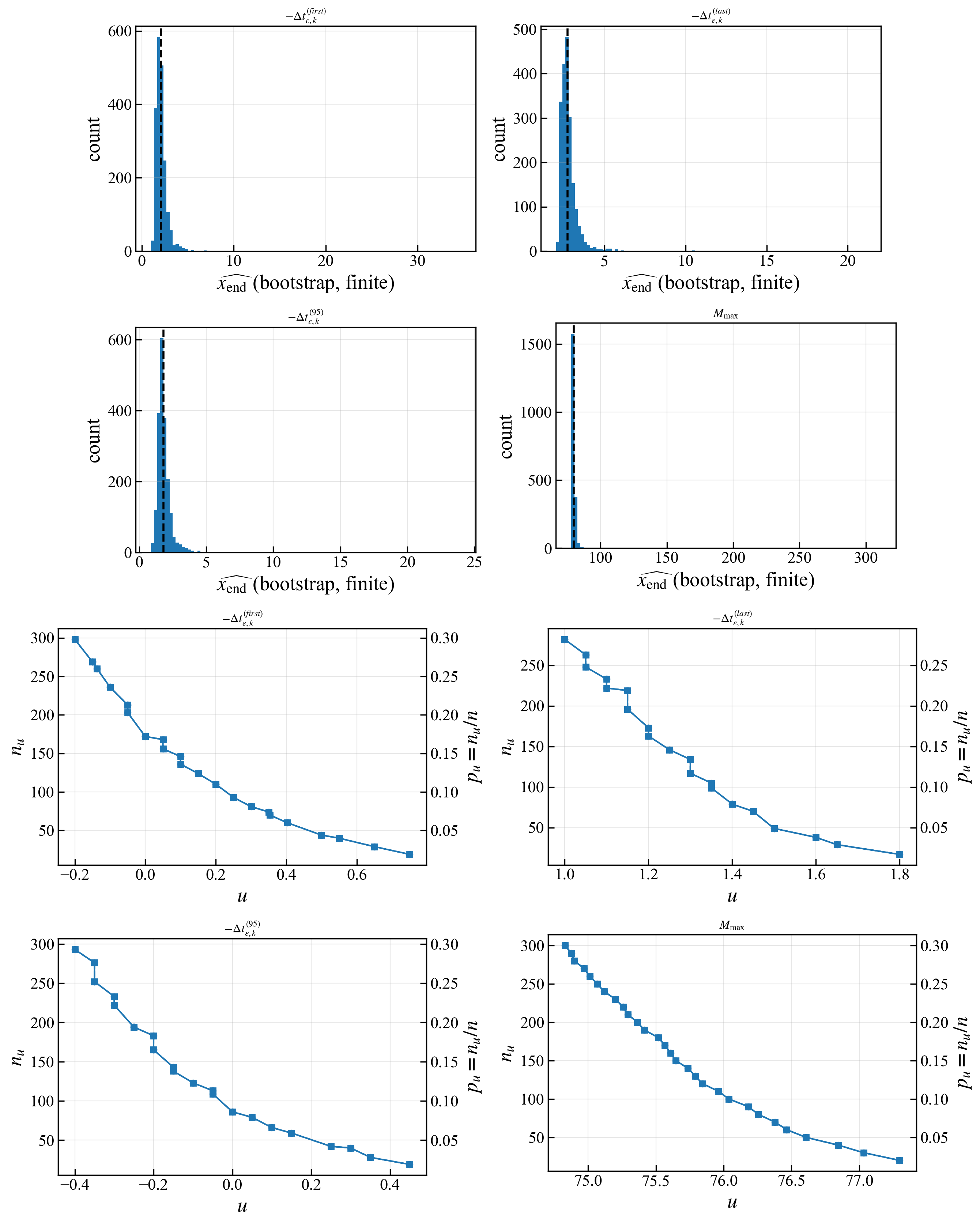}
  \caption{Bootstrap and threshold tradeoff diagnostics for fixed-threshold POT fits.
  (a)--(d) Bootstrap histograms of the finite endpoint estimate $\hat x_{\mathrm{end}}$ (computed only over replicates with $\hat\xi<0$), with the MLE from the original sample overlaid (vertical dashed line).
  (e)--(h) Exceedance count $n_u$ and exceedance fraction $p_u=n_u/n$ as functions of the threshold $u$, illustrating the basic tradeoff between tail purity (higher $u$) and available tail sample size (smaller $n_u$).}
  \label{fig:fig7}
\end{figure*}

To address potential sensitivity to the threshold choice $u$ and to quantify uncertainty in the generalized Pareto (GPD) tail parameters, we carried out complementary diagnostic checks. First, as $u$ is increased, the number of exceedances $n_u$ must decrease by construction; the resulting threshold--sample-size tradeoff is shown explicitly in Fig.~\ref{fig:fig7}(e--h). Second, we assess finite-sample uncertainty by fixed-threshold bootstrap resampling ($B=2000$), summarized in Table~\ref{tab:bootstrap_ci}.
Here, ``finite endpoint'' statistics are computed only over bootstrap replicates with $\hat\xi<0$, while $p_\infty$ denotes the bootstrap frequency of $\hat\xi\ge 0$ (unbounded endpoint).

\begin{figure*}[t]
  \centering
  \includegraphics[width=0.8\linewidth]{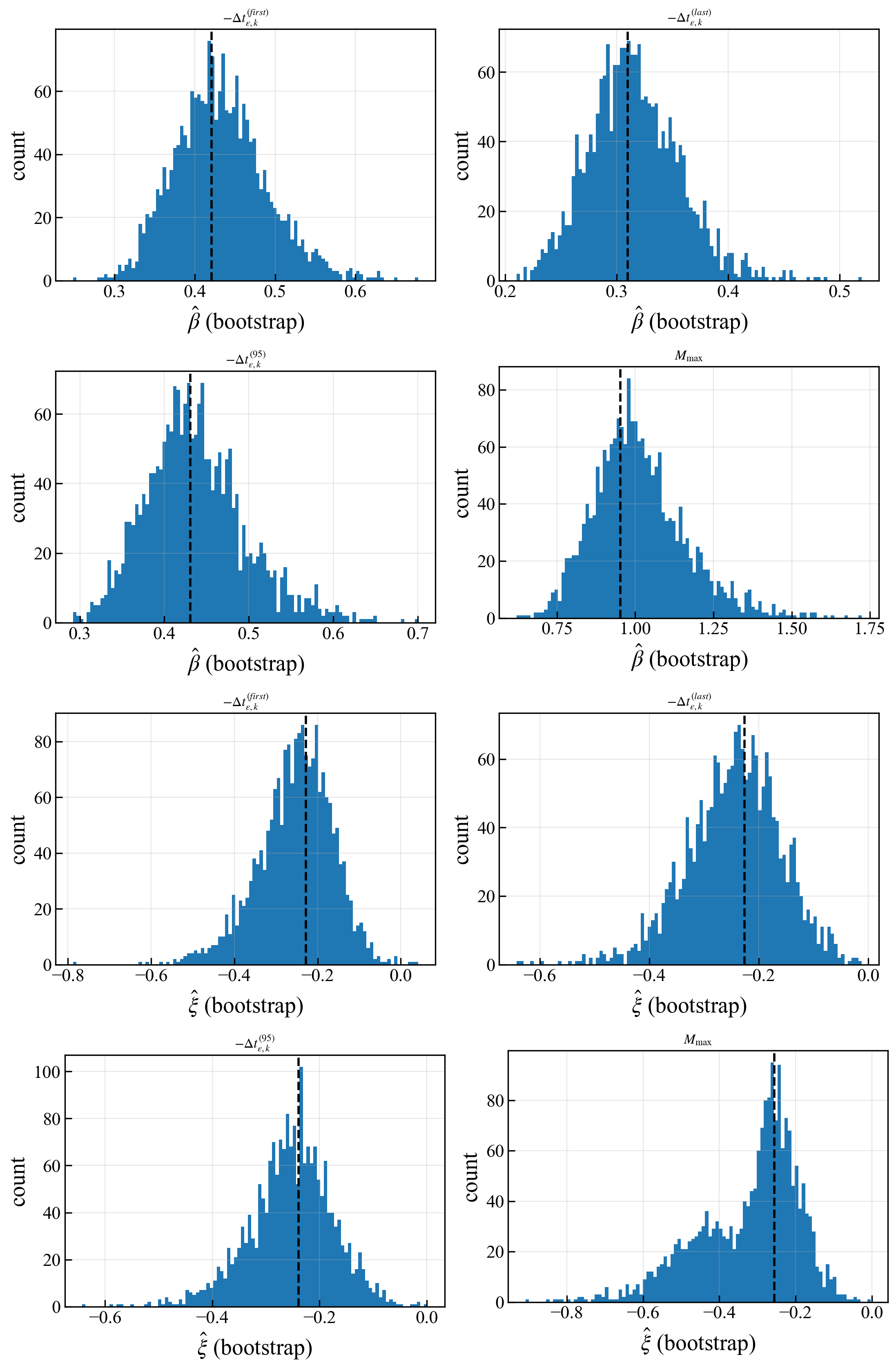}
  \caption{Bootstrap histograms of the GPD scale $\hat\beta$ and shape $\hat\xi$ for fixed-threshold POT fits ($B=2000$).
  Vertical dashed lines indicate the MLEs from the original sample.
  While endpoint estimates concentrate tightly (Fig.~\ref{fig:fig7}), the parameter histograms can exhibit skewness and occasional secondary modes, reflecting the flexibility of the GPD family and finite-sample sensitivity of $(\xi,\beta)$ under a bounded-tail constraint.}
  \label{fig:fig8}
\end{figure*}

In addition to the endpoint resampling shown in Fig.~\ref{fig:fig7}, we also inspect the bootstrap variability of the GPD parameters themselves.
Figure~\ref{fig:fig8} displays the fixed-threshold bootstrap distributions ($B=2000$) of the fitted shape $\hat{\xi}$ and scale $\hat{\beta}$ (dashed lines: original-sample MLEs).
Even when the implied endpoint estimate is relatively concentrated, the marginal histograms of $(\hat{\xi},\hat{\beta})$ can be noticeably asymmetric and may occasionally exhibit secondary structure.
This behavior is consistent with the strong finite-sample dependence between $\xi$ and $\beta$ in POT inference and with the fact that the endpoint $x_{\mathrm{end}} = u - \beta/\xi$ (for $\xi<0$) is a nonlinear combination of the two parameters; different $(\xi,\beta)$ pairs can therefore yield comparable endpoint values.
\HLF{The bootstrap mass remains overwhelmingly in the $\hat{\xi}<0$ regime (see also the small unbounded-endpoint frequency $p_{\infty}$ in Table~\ref{tab:bootstrap_ci}), supporting the bounded-tail interpretation at the adopted threshold.}

\begin{figure*}[t]
  \centering
  \includegraphics[width=\linewidth]{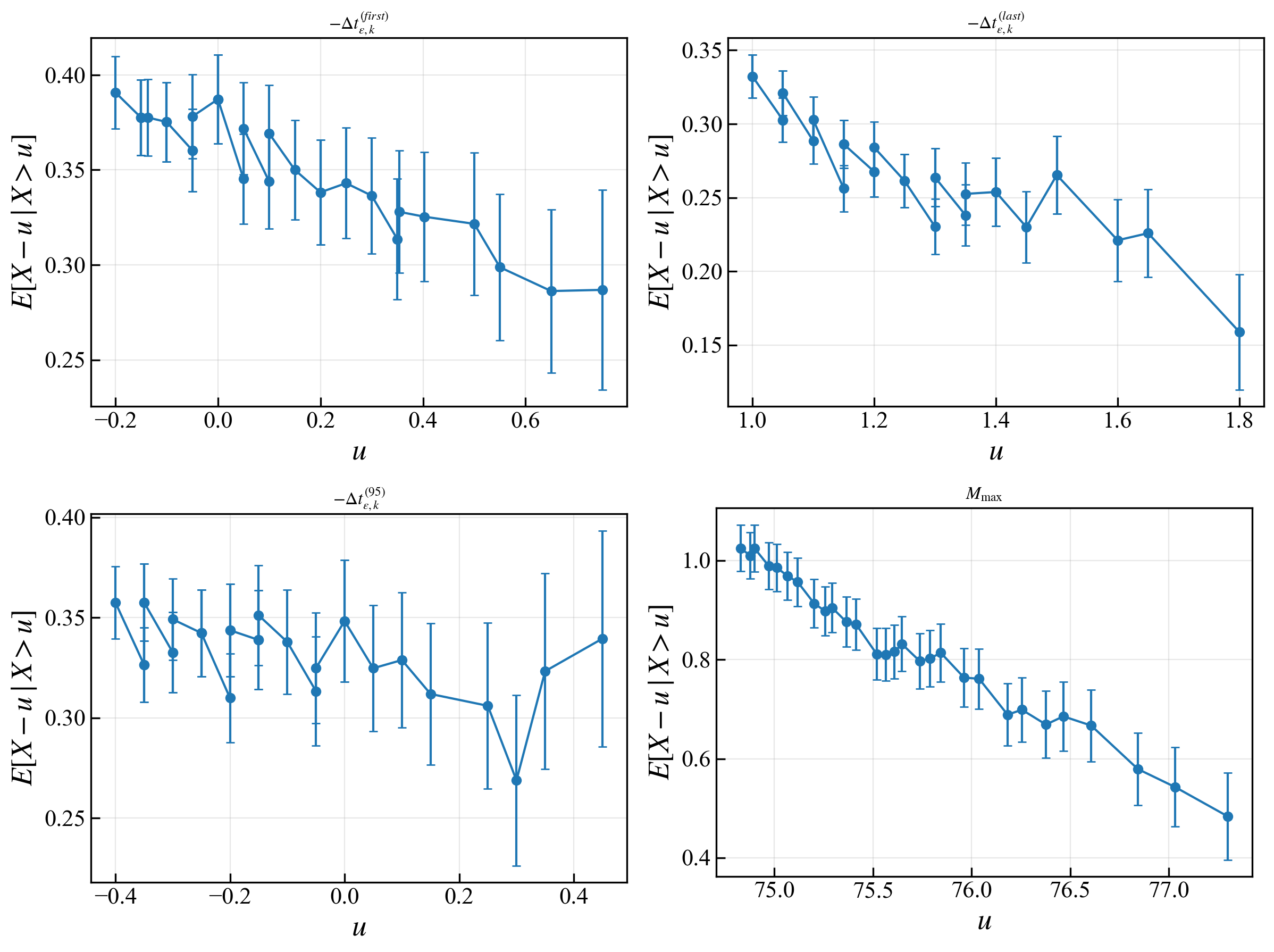}
  \caption{Mean residual life (MRL) diagnostics, showing $\mathbb{E}[X-u\mid X>u]$ versus threshold $u$ (with error bars) for each analyzed variable. Error bars indicate $\pm 1$ standard error of the mean excess computed from exceedances above $u$.
  For data well described by a GPD above $u$, the MRL is approximately linear in $u$;
  the present plots provide a qualitative check, but become noisy at high thresholds as $n_u$ decreases.}
  \label{fig:fig9}
\end{figure*}

\begin{figure*}[t]
  \centering
  \includegraphics[width=\linewidth]{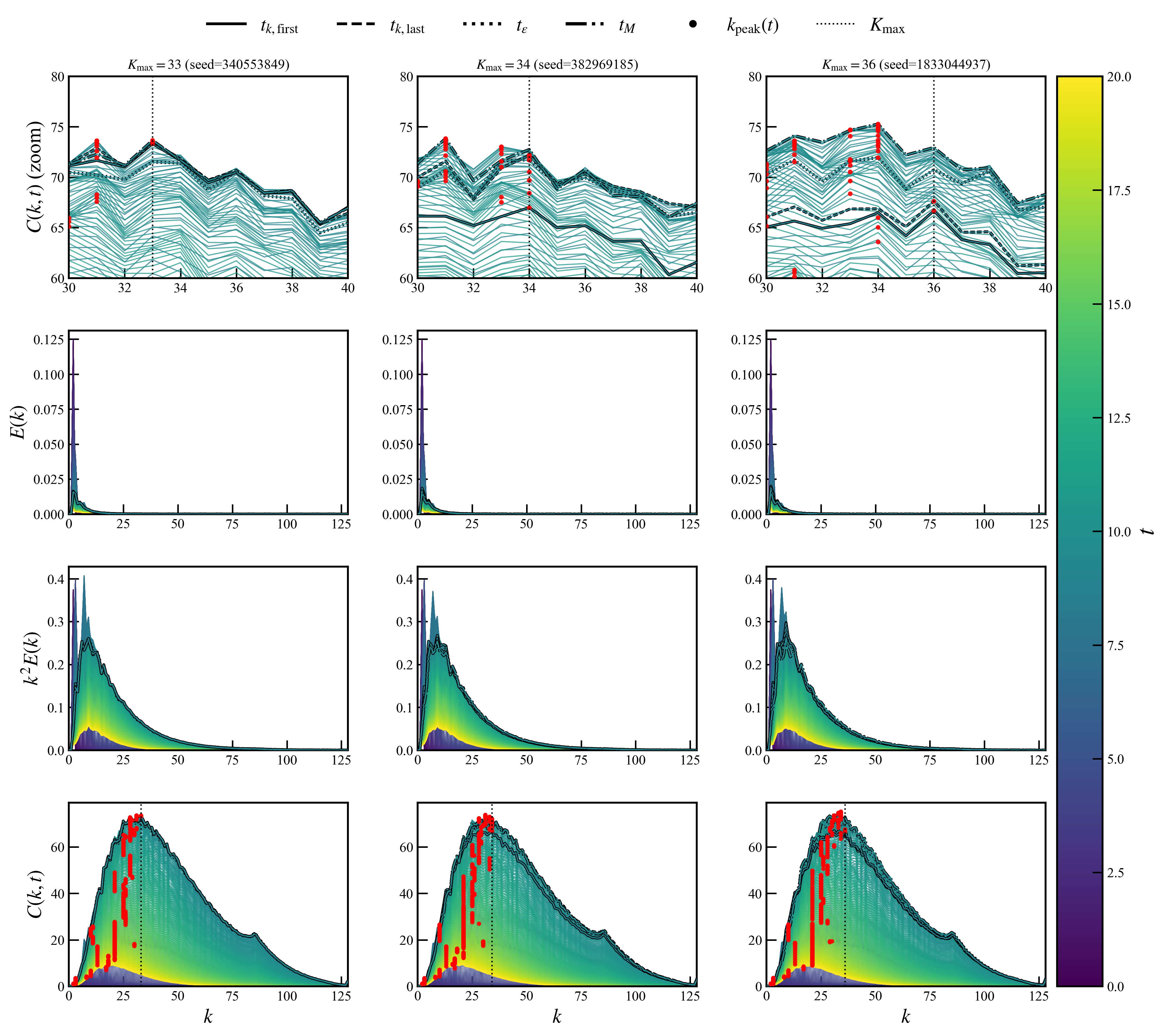}
\caption{
Spectral evolution for three representative perturbed Taylor--Green vortex (TGV) realizations
($N=256^3$, $\nu=10^{-3}$) illustrating the discrete $K_{\max}$ branches.
Columns correspond to runs with $K_{\max}=33$, 34, and 36 (seeds are given in the panel titles).
All panels show shell-averaged spectra colored by time $t$ (shared color bar).
The top row shows a zoomed view ($k\in[30,40]$) of the curl-of-vorticity spectrum $C(k,t)$,
defined as the shell-averaged spectrum of $|\nabla\times\boldsymbol{\omega}|^2$.
Rows 2--4 show, respectively, $E(k,t)$, $k^2E(k,t)$, and $C(k,t)$ over the full resolved wavenumber range.
Red markers indicate the instantaneous peak wavenumber
$k_{\mathrm{peak}}(t)=\arg\max_k C(k,t)$, and the vertical dotted line marks the run-wise maximum
$K_{\max}=\max_t k_{\mathrm{peak}}(t)$.
Thick black curves highlight spectra at $t_{k,\mathrm{first}}$ (solid), $t_{k,\mathrm{last}}$ (dashed),
$t_{\varepsilon}$ (dotted), and $t_M$ (dash--dotted), where
$M(t)=\max_k C(k,t)$ and $t_M=\arg\max_t M(t)$.
\HLF{Here $k_{\mathrm{peak}}(t_M)$ does not necessarily coincide with $K_{\max}$:
$K_{\max}$ is the largest peak wavenumber attained over the trajectory.}
}
\label{fig:fig10}
\end{figure*}

As a complementary threshold diagnostic, Fig.~\ref{fig:fig9} shows mean residual life (MRL) curves, i.e. the empirical mean excess $e(u)=\mathbb{E}[X-u\,|\,X>u]$ plotted against the threshold $u$ with $\pm 1$ standard-error bars.
For a GPD tail, $e(u)$ is approximately affine in $u$; in particular, when $\xi<0$ it is expected to decrease roughly linearly as $u$ approaches the finite endpoint.
The present MRL plots are broadly consistent with this expectation over intermediate thresholds, whereas they become progressively noisier at the highest thresholds because the number of exceedances $n_u$ rapidly decreases (cf.\ Fig.~\ref{fig:fig7}(e--h)). Taken together, Figs.~\ref{fig:fig7}--\ref{fig:fig9} provide a qualitative validation of the chosen POT threshold and a transparent view of the finite-sample uncertainty in the inferred tail parameters.

\begin{figure*}[t]
  \centering
  \includegraphics[width=\linewidth]{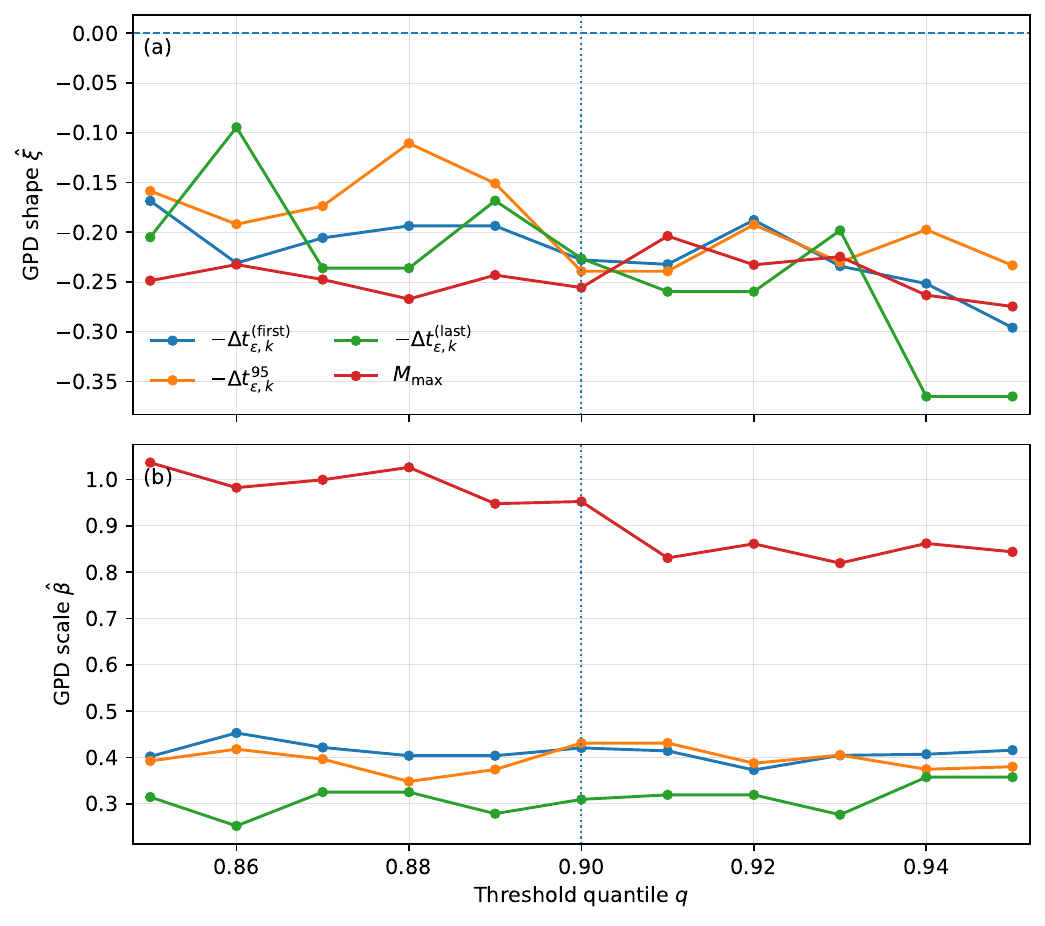}
  \caption{\HLS{Threshold-stability diagnostics for the GPD fits. The empirical threshold quantile is varied over $q\in[0.85,0.95]$. Panel (a) shows the fitted shape parameter $\hat{\xi}$, and panel (b) shows the fitted scale parameter $\hat{\beta}$. The vertical dotted line marks the threshold used in the main POT table, $q=0.9$, and the horizontal dashed line in panel (a) marks $\hat{\xi}=0$.}}
  \label{fig:fig11}
\end{figure*}

\begin{figure*}[t]
  \centering
  \includegraphics[width=\linewidth]{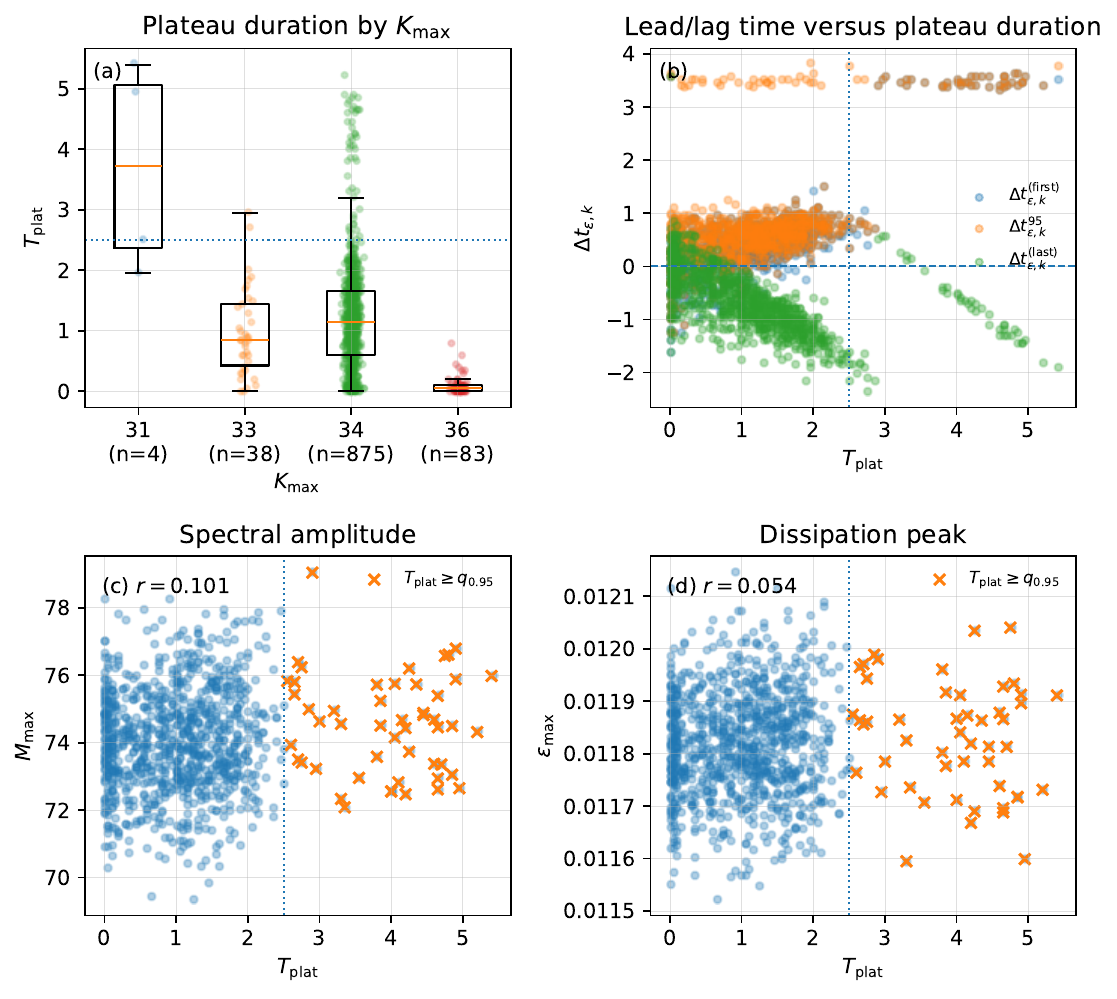}
  \caption{\HLS{Plateau-duration diagnostics for the resolved reference ensemble. Panel (a) shows the distribution of $T_{\mathrm{plat}}$ grouped by $K_{\max}$. Panel (b) compares $T_{\mathrm{plat}}$ with the lead--lag variables for the three timing definitions. Panels (c) and (d) compare $T_{\mathrm{plat}}$ with $M_{\max}$ and $\varepsilon_{\max}$, respectively. The dotted vertical or horizontal line marks the upper $5\%$ plateau-duration threshold $T_{\mathrm{plat}}=2.5025$.}}
  \label{fig:fig12}
\end{figure*}

\HLS{To assess threshold stability more directly, we repeated the GPD fits for empirical quantile thresholds $q\in[0.85,0.95]$. Figure~\ref{fig:fig11} shows the resulting $\hat{\xi}(q)$ and $\hat{\beta}(q)$ for $-\Delta t_{\varepsilon,k}^{(\mathrm{first})}$, $-\Delta t_{\varepsilon,k}^{95}$, $-\Delta t_{\varepsilon,k}^{(\mathrm{last})}$, and $M_{\max}$. The fitted shape parameter remains negative over the inspected range for all four variables. The estimates fluctuate more strongly near the highest thresholds because the number of exceedances decreases, but no sign change of $\hat{\xi}$ is observed. The corresponding scale parameter varies smoothly over intermediate thresholds. This threshold scan supports the use of $q=0.9$ as a compromise between tail selectivity and finite-sample variability, while also reinforcing the interpretation that the inferred bounded-tail behavior is an effective, finite-sample property of the present ensemble rather than a continuum-level invariant.}

\HLT{The threshold scan in Fig.~\ref{fig:fig11} is used as a stability diagnostic rather than as a complete propagation of threshold-selection uncertainty. It shows that the sign of $\hat{\xi}$ is stable over the inspected range, while the finite endpoint and parameter values remain threshold dependent and should be interpreted with the finite-sample caution described above. Endpoint estimates require particular caution because $x_{\mathrm{end}}=u-\beta/\xi$ is a nonlinear function of the fitted parameters and is sensitive when $\hat\xi$ is only mildly negative. In the present work, $x_{\mathrm{end}}$ is used as a diagnostic estimate of the largest plausible lag or amplitude under the adopted finite-resolution protocol, not as a sharply determined physical upper bound.}

\begin{table*}[t]
\caption{Bootstrap uncertainty summary for fixed-threshold POT fits ($B=2000$). 
Here ``finite endpoint'' statistics are computed only over bootstrap replicates with $\hat{\xi}<0$.
The fraction $p_{\infty}$ denotes the bootstrap frequency of $\hat{\xi}\ge 0$ (unbounded endpoint).
For each variable, the reported $\hat{\xi}$, $\hat{\beta}$, and $\hat{x}_{\mathrm{end}}$ are maximum-likelihood estimates (MLEs), with 95\% confidence intervals given in brackets.}
\label{tab:bootstrap_ci}
\begin{ruledtabular}
\begin{tabular}{lrrrr}
Variable & $\hat{\xi}$ (95\% CI) & $\hat{\beta}$ (95\% CI) & $\hat{x}_{\mathrm{end}}$ (95\% CI) & $p_{\infty}$ \\
\midrule
$-\Delta t_{\varepsilon,k}^{\mathrm{(first)}}$
& $-0.228\ [ -0.451,\ -0.097]$
& $0.421\ [0.332,\ 0.557]$
& $2.097\ [1.375,\ 3.886]$
& $0.002$ \\
$-\Delta t_{\varepsilon,k}^{(s=0.95)}$
& $-0.239\ [ -0.424,\ -0.110]$
& $0.431\ [0.336,\ 0.578]$
& $1.803\ [1.183,\ 3.316]$
& $0.001$ \\
$-\Delta t_{\varepsilon,k}^{\mathrm{(last)}}$
& $-0.226\ [ -0.423,\ -0.086]$
& $0.310\ [0.241,\ 0.403]$
& $2.718\ [2.237,\ 4.264]$
& $0.000$ \\
$M_{\max}$
& $-0.256\ [ -0.622,\ -0.125]$
& $0.953\ [0.769,\ 1.363]$
& $79.764\ [78.039,\ 82.468]$
& $0.000$ \\
\end{tabular}
\end{ruledtabular}
\end{table*}

\bibliography{main}

\end{document}